\documentclass[12pt,preprint]{aastex}
\def\degree{\ifmmode {^\circ}\else {$^\circ$}\fi}
\def\rstar{\ifmmode {\, R_{\star}}\else $R_{\star}$\fi}
\def\msol{\ifmmode {\, M_{\odot}}\else $M_{\odot}$\fi}
\def\rsol{\ifmmode {\, R_{\odot}}\else $R_{\odot}$\fi}
\def\lsol{\ifmmode {\, L_{\odot}}\else $L_{\odot}$\fi}
\def\msolyr{\ifmmode {\,M_{\odot}\,{\rm yr}^{-1}}\else $M_{\odot}\,{\rm yr}^{-
1}$\fi}
\def\mdot{\ifmmode {\,\dot{M}}\else $\dot{M}$\fi}
\def\mdotyr{\ifmmode {\,\dot{M}\,yr^{-1}}\else $\dot{M}\,yr^{-1}$\fi}

\begin{document}

\title{The MACHO Project LMC Variable Star Inventory: X.\\ The R Coronae Borealis 
Stars} 
\author {C. Alcock$^{1,2}$, R.A. Allsman$^3$, D.R. Alves$^{4}$,
T.S. Axelrod$^5$, A. Becker$^{6}$, D.P. Bennett$^{1,7}$, Geoffrey C. 
Clayton$^{8,9}$, 
K.H. Cook$^{1,2,9}$,  N. Dalal$^{2,10}$, A.J. Drake$^{1,5}$,
K.C. Freeman$^5$, M. Geha$^{1}$, K.D. Gordon$^{9,11}$, K. Griest$^{2,10}$, D. Kilkenny$^{12}$, M.J. 
Lehner$^{13}$, S.L. Marshall$^{1,2}$, D. Minniti$^{1,14}$, K.A. Misselt$^{15}$, C.A. Nelson$^{1,16}$,
B.A. Peterson$^5$, P. Popowski$^1$, M.R. Pratt$^{6}$, P.J. Quinn$^{17}$, C.W. 
Stubbs$^{2,5,6}$, W. Sutherland$^{18}$, A. Tomaney$^6$, T. Vandehei$^{2,9}$, and 
D.L. Welch$^{19}$, (The MACHO Collaboration)}

\slugcomment{Scientific correspondence should be directed to Geoffrey C. Clayton}
\altaffiltext{1}{Lawrence Livermore National Laboratory, Livermore, CA 94550; alcock@igpp.llnl.gov,
kcook@igpp.llnl.gov, adrake@igpp.llnl.gov, mgeha@igpp.llnl.gov, stuart@igpp.llnl.gov,
dminniti@igpp.llnl.gov, cnelson@igpp.llnl.gov, popowski@igpp.llnl.gov}
\altaffiltext{2}{Center for Particle Astrophysics, University of California, Berkeley, CA 94720}
\altaffiltext{3}{Supercomputing Facility, Australian National University, Canberra, ACT 0200, Australia;
Robyn.Allsman@anu.edu.au}
\altaffiltext{4}{Space Telescope Science Institute, 3700 San Martin Drive, Baltimore, MD 21218; 
alves@stsci.edu}
\altaffiltext{5}{Research School of Astronomy and Astrophysics, Canberra, Weston Creek, ACT 2611, 
Australia;
tsa@mso.anu.edu.au, kcf@mso.anu.edu.au, peterson@mso.anu.edu.au}
\altaffiltext{6}{Departments of Astronomy and Physics, University of Washington, Seattle, WA 98195;
becker@astro.washington.edu, stubbs@astro.washington.edu}
\altaffiltext{7}{Department of Physics, University of Notre Dame, Notre Dame, IN 46556;
bennett@bustard.phys.nd.edu}
\altaffiltext{8}{Dept. of Physics \& Astronomy, Louisiana State University, Baton Rouge, LA 70803; 
gclayton@fenway.phys.lsu.edu}
\altaffiltext{9}{Visiting Astronomer, Cerro Tololo Inter-American Observatory} 
\altaffiltext{10}{Department of Physics, University of California at San Diego, San Diego, CA 92093;
endall@physics.ucsd.edu, kgriest@ucsd.edu, vandehei@astrophys.ucsd.edu}
\altaffiltext{11}{Steward Observatory, University of Arizona, Tucson, AZ 85721; kgordon@as.arizona.edu}
\altaffiltext{12} {South African Astronomical Observatory, P.O. Box 9, Observatory 7935, 
South Africa; dmk@da.saao.ac.za}
\altaffiltext{13}{Department of Physics, University of Sheffield, Sheffield S3 7RH, England, UK;
m.lehner@sheffield.ac.uk}
\altaffiltext{14}{Departimento de Astronomia, P. Universidad Catolica, Casilla 104, Santiago 22, Chile;
dante@astro.puc.cl} 
\altaffiltext{15}{Infrared Astrophysics Branch, Code 685, NASA/Goddard Space Flight Center, Greenbelt, MD 20771; 
misselt@idlastro.gsfc.nasa.gov}
\altaffiltext{16}{Department of Physics, University of California at Berkeley, Berkeley, CA 94720} 
\altaffiltext{17}{European Southern Observatory, Karl Schwarzchild Strasse 2, D-8574 8 G\"{a}rching bei 
M\"{u}nchen,
Germany; pjq@eso.org}
\altaffiltext{18}{Department of Physics, University of Oxford, Oxford OX1 3RH, England, UK;
w.sutherland@physics.ox.ac.uk}
\altaffiltext{19}{McMaster University, Hamilton, Ontario Canada L8S 4M1; welch@physics.mcmaster.ca}

\begin{abstract}
We report the discovery of eight new R Coronae 
Borealis (RCB) stars in the Large Magellanic Cloud (LMC) using the MACHO project 
photometry database.  The 
discovery of these new stars increases the number of known RCB stars in the LMC to thirteen. 
We have also discovered four stars
similar to the Galactic variable DY Per. These stars decline much more slowly and are 
cooler than the RCB stars.
The absolute luminosities of the Galactic RCB  stars are unknown since 
there is no direct measurement of the distance to any Galactic RCB 
star. Hence, the importance of the LMC RCB stars. We find a much larger range of absolute 
magnitudes (M$_V$ = -2.5 to -5 mag) than inferred from the
small pre-MACHO sample of LMC RCB stars. It is likely that there is a temperature - M$_V$ 
relationship with the cooler
stars being intrinsically fainter. Cool ($\sim$ 5000 K) RCB stars are much more common 
than previously thought based on the 
Galactic RCB star sample.  Using the fairly complete sample of RCB stars discovered in the MACHO fields, 
we have 
estimated the likely number of RCB stars in the Galaxy to be $\sim$3,200.
The SMC MACHO fields were also searched for RCB stars but none were found. 
\end{abstract}

% Keywords should be included, but they are not printed in the hardcopy.


\section{Introduction}
The R~Coronae Borealis (RCB) stars are rare, hydrogen-deficient, carbon-rich 
supergiants which undergo spectacular declines in brightness of up to 8 
magnitudes at irregular intervals as dust forms along the line 
of sight (Clayton 1996). 
Their rarity may stem from the fact that they are in an extremely rapid 
phase of evolution toward white dwarfs. 
So understanding the RCB stars is a key test
for any theory which aims to explain hydrogen deficiency in
post-Asymptotic Giant Branch (AGB) stars.  
AGB and post-AGB stars are the dominant formation sites for refractory
grains subsequently injected into the interstellar medium, and therefore
they play an important role in any theory of dust condensation.

There are two major evolutionary models for the origin of RCB stars: 
the Double Degenerate and the Final Helium Shell Flash (Iben, Tutukov, \& Yungleson 1996a). 
The former involves the merger of two white dwarfs, and the latter involves a 
white dwarf/evolved Planetary Nebula (PN) central star which
is blown up to supergiant size by a final helium shell flash.
In the final flash model, there is a close relationship between RCB stars and
PN. The connection between RCB stars and PN has recently become stronger, since 
the central stars of three old PN's (Sakurai's Object, V605 Aql and FG Sge) 
have had observed outbursts that transformed them from hot evolved central 
stars into cool giants with the spectral properties of an RCB star
(Kerber et al. 1999; Asplund et al. 1999; Clayton \& De Marco 1997; 
Gonzalez et al. 1998).  

However, the absolute luminosities of the RCB  stars are unknown. 
There is no direct measurement of the distance to any Galactic RCB 
star (Alcock et al. 1996 and references therein). 
A group of RCB stars were measured by HIPPARCOS but the data provided only 
lower limits on their distances  (Cottrell \& Lawson 1998; Trimble \& Kundu 1997).
So the only source
of distances and absolute luminosities for the RCB stars is the LMC.
The distance to the LMC is fairly well determined, m-M = 18.4 mag (e.g., Nelson et al. 2000).
Until recently, only three RCB stars were known in the LMC (Feast 1972). On the basis of 
this small sample of stars, an absolute magnitude range of $M_V$ = -4 to -5 is inferred.
Alcock et al. (1996) reported the discovery of two additional LMC RCB stars.  
These two stars were both fainter, $M_V \sim$ -3.5 but one of the two is a member of the
subclass of hot RCB stars and may not be comparable to the cooler stars. 
 
This paper reports the results of a more extensive search of the MACHO database for new RCB stars.

\section{Observational Data}
\subsection{MACHO Photometry}
The MACHO Project (Alcock et al. 1992) is an astronomical survey experiment designed to 
obtain multi-epoch, two-color CCD photometry of millions of stars in the LMC (also, the 
Galactic bulge and SMC).  The survey makes use of a dedicated 1.27m telescope at Mount 
Stromlo, Australia and because of its southerly latitude is able to obtain observations of the 
LMC year round (Hart et al. 1996). The camera built specifically for this project (Stubbs et 
al. 1993) has a 
field of view of 0.5 square degrees which is achieved by imaging at prime focus. 
Observations are obtained in two bandpasses simultaneously, using a dichroic beamsplitter 
to direct the ``blue" ($\sim$4400-5900 \AA) and ``red" ($\sim$5900-7800 \AA) light onto 
2x2 mosaics of 2048x2048 Loral CCD's. These bandpasses are referred to as 
$V_{MACHO}$ and $R_{MACHO}$, respectively. Images are obtained and read out 
simultaneously. The 15 \micron~pixel size maps to $0\farcs63$~on the sky. The data were 
reduced using a profile-fitting photometry routine known as SODOPHOT, derived from 
DoPHOT (Schecter, Mateo, \& Saha 1993). This implementation employs a single starlist 
generated from frames obtained in good seeing.
The results reported in this survey comprise only a fraction of the planned data acquisition 
of the MACHO project. 
The MACHO data were acquired for 82 LMC fields
covering approximately 40 square degrees and were monitored for about 7 years from 1992 to 1999. 
Most of the data come from the top-22 fields which contain
approximately 9 million stars (Alcock et al. 1997a, 1999). 
These data have been searched for variable stars 
and microlensing candidates and over 40,000 variables have been found, most newly 
discovered. The great majority of these fall into four well known classes: there are 
approximately 25,000 very red semi-regular or irregular variables, 1500 Cepheids, 8000 
RR 
Lyraes, and 1200 eclipsing binaries (Cook et al. 1995). Typically, the dataset for a given 
star 
covers a timespan of about 2700 days and contains up to 1500 photometric measurements 
(multiple observations are obtained on a given night whenever conditions allow).  The 
output photometry contains flags indicating suspicion of errors due to crowding, seeing, 
array defects, and
radiation events. 

The MACHO lightcurves for the RCB candidate stars, including all available data, are shown in Figure 1.
Tables containing all the MACHO photometric data for these stars are given in the Appendix.
HV 12842 lies outside the MACHO fields and so has no lightcurve.
Only data free from suspected errors are plotted and included in the tables. 
Typical photometric uncertainties are in the range 1.5-2~\%. The 
$V_{MACHO}$ and $R_{MACHO}$ bandpasses have been converted to 
Kron-Cousins (KC) V and R bandpasses using transformations determined from 
the internal calibrations of the MACHO database (Alcock et al. 1999).
Three stars in the sample, 16.6541.22, 20.5036.12, and 21.7407.7 lie outside
the top-22 fields which
have been photometrically calibrated. The V and R magnitudes for these fields were 
obtained by bootstrapping from the 
top-22-field
calibration and should be viewed with caution. Also, 18.3325.148, 20.5036.12, and 21.7407.7 
are the three brightest stars in
the sample and suffer from possible saturation problems. In particular, the MACHO V-R colors 
for 21.7407.7 are $\sim$-0.2 
but are actually $\sim$0.2 (Goldsmith et al. 1990).
The $(V-R)_{KC}$ colors for each star are also plotted in Figure 1. 
The stars are listed in Table 1. Two designations are given for each MACHO star, the
MACHO name which includes the position of the star, and the standard field.tile.sequence number which 
refers to a particular star in the database.
Finding charts for the stars are shown in Figure 2.

\subsection{JHK Photometry}
The Cerro Tololo Infrared Imager (CIRIM) uses a 256x256 HgCdTe NICMOS 3 array.
There were two runs in 1996 and 1999 on the CTIO 1.5m telescope. The data are listed in Table 2.
The typical 1-$\sigma$ errors are 0.02-0.03 mag in the J-band and 0.04-0.05 mag in the H- and K-bands.
None of the data from the 1999 run are included because of their large uncertainties.
In addition, 2MASS photometry is now available for most of the sample. 
These data are also listed in Table 2. The typical 1-$\sigma$ 2MASS errors 
are 0.03 mag in all three bands. The Julian Dates of the observations are also listed in Table 2.

\subsection{Spectroscopic Data}
Spectroscopic observations were obtained from 1995 to 1998.
In 1995-1996, the spectra were obtained with the Reticon photon-counting system
on the image-tube spectrograph on the SAAO 1.9m telescope at Sutherland, South Africa.
The grating used gives a reciprocal dispersion of 100 $\AA~mm^{-1}$ and a
resolution of approximately 4 \AA, giving a useful range of about
3600-5200 \AA~at the angle setting used.  The spectrograph is a two-aperture
instrument recording the star and sky simultaneously.  Normal operating procedure is to 
measure
the star through one aperture and then the other, so the sequence
goes arc, star in A, arc, star in B, arc. Each star is then wavelength
calibrated by the two arcs on either side and the results of star in
A and B are added together after flat-field correction and sky subtraction.
Flux calibration is done by observing one
standard star each night.  
The fluxes are given in $erg~cm^{-2}~s^{-1}~\AA^{-1}$.
In 1997-1998, the spectrograph was adapted to take a SITe CCD chip. The observations 
were taken using a grating with a reciprocal dispersion of 210 $\AA~mm^{-1}$ and a
resolution of approximately 5 \AA, giving a useful range of about
3500-7600 \AA~at the angle setting used.

Spectra have been obtained for all the stars in Table 1 except 6.6575.13 and 16.5641.22. A spectrum of the latter star is 
shown in Bessell \& Wood (1983). No spectrum exists for 6.6575.13 since it has been continuously in decline since 1993. 
A spectrum of the hot RCB star, 11.8632.2507, is shown in Alcock et al. (1996). Further spectra are shown in Clayton et al.
(2001).
The spectra of the remaining stars are shown in Figure 3. All of the spectra were obtained when the stars were at or near
maximum light.
These spectra are sums of all individual scans.

\section{New RCB Stars in the LMC}
From the night in 1795 when Edward Pigott first noticed that R CrB had apparently disappeared from the sky,
RCB stars have been characterized primarily on the basis of their lightcurves. They are the only 
intrinsic variables that undergo sudden and severe drops in brightness from their light maximum at irregular 
intervals. 
However, this definition has resulted in many irregular variables with poor lightcurve coverage being identified
as RCB stars (Payne-Gaposchkin \& Gaposchkin 1938).  Spectroscopic confirmation has been used to weed out non-RCB 
stars from the class.  Many turn out to be symbiotic, cataclysmic or semi-regular variables (Lawson \& Cottrell 
1990a).  Much of this sorely needed spectroscopic work has been performed by Kilkenny and collaborators 
(Clayton 1996 and references therein). However, if a well sampled lightcurve is available then an identification with
the RCB class may be made with fairly high confidence because of the  
distinctive nature of the RCB declines. Following the definition of Payne-Gaposchkin \& Gaposchkin (1938), a typical 
RCB lightcurve has:\\
$\bullet$ uniform brightness at maximum which may last for months or years.\\
$\bullet$ A sudden drop in brightness of more than three magnitudes taking a few days or weeks.\\
$\bullet$ Recovery to maximum light, which  is typically slower, taking months or years.\\

When spectra are obtained, we see in addition that the typical RCB star has:\\
$\bullet$ Weak or absent hydrogen lines and molecular bands (CH). 
\footnote{V854 Cen, one of the most active RCB stars, shows fairly strong Balmer lines and CH band 
(Kilkenny \& Marang
1989; Lawson \& Cottrell 1989).}\\
$\bullet$ Strong carbon lines and molecular bands (CN, $C_2$).\\
$\bullet$ Little or no $^{13}C$.\\

Other observables of the typical RCB star are:\\
$\bullet$ Regular or semi-regular pulsations with $\Delta$V of a few tenths of a magnitude and periods of 
40-100 days.\\
$\bullet$ An infrared excess.\\
$\bullet$ Effective temperature between 5000 and 7000 K. A small subclass is much hotter with effective temperatures
of about 20,000 K.  

The first RCB star to be discovered in the LMC was HV 966 (W Men, 21.7407.7) (Luyten 1927). He identified it as an 
RCB star on the basis of its irregular and sudden dips in brightness. Much 
later, spectra confirming 
its resemblance to R CrB and its membership in the LMC were obtained (Feast 1956; Rodgers 1970; Feast 1972). 
Subsequently, two other stars, HV 5637 
(Hodge, \& Wright 1969)
and HV 12842 (Payne-Gaposchkin 1971), were identified as members of the RCB class on the basis of their lightcurves. 
Confirming spectra were soon obtained of these stars (Feast 1972).
A fourth star, HV 12671, was listed by Payne-Gaposchkin (1971) as an RCB star. It is now thought to be a 
carbon-symbiotic star (Allen 1980; Lawson et al. 1990). Two new RCB stars were previously discovered using the 
MACHO database, HV 2671 (11.8632.2507) and 81.8394.1358 (Alcock et al. 1996).

The final database of MACHO variables for the top-22 fields was searched for stars which underwent large sudden 
brightness variations.  Candidates were selected from the thousands of variable star lightcurves by
selecting out those with large amplitude variations that were not periodic. 
These lightcurves were then viewed 
by eye and candidates were selected as having distinctive RCB lightcurve behavior. 

In addition, the MACHO lightcurves for stars listed as irregular variables of large range by
Payne-Gaposchkin (1971) were
examined. HV 942 (6.6696.60) and HV 12524 (18.3325.148) were found to be RCB stars. Stars listed by Hughes (1989)
as having RCB-type lightcurves were also checked. None of those lying in MACHO fields 
turned out to be RCB stars. However, three other 
variables in the Hughes list are members of the MACHO RCB star sample.
Of the stars in our sample, three appear in the list of carbon stars in the LMC compiled by 
Sanduleak \& Phillip (1977). However, none
appear in the carbon star catalogs of Westerlund et al. (1978) or Blanco \& McCarthy (1990).
 
In addition to the five RCB stars already known in the LMC, eight stars (6.6575.13, 6.6696.60, 12.10803.56, 16.5641.22,
18.3325.148, 
79.5743.15, 80.6956.207, and 80.7559.28) which show lightcurves with deep, sharp declines are clearly RCB stars. See
Figure 1. 
Two stars, 6.6575.13 and 6.6696.60 were independently discovered to be RCB stars by Wood \& Cohen (2001).
As summarized in Table 3, these stars share most or all of the photometric and spectroscopic criteria listed above
for the typical RCB star. In particular, with the exception of 18.3325.148, as shown in Figure 1, 
all of these stars show the unique 
sharp, deep, irregular declines characteristic of RCB stars. 
This has been quantified in Table 3 as dm/dt, the number of magnitudes per day that the star fades. 
The lightcurve for 18.3325.148 shows only the long recovery from a decline often seen in Galactic RCB stars.
Subsequently, spectra were obtained for all except 6.6575.13 showing
that they are indeed RCB stars. The star, 6.6575.13, entered a very deep decline early in the MACHO era 
and has never recovered enough for a
spectrum to be obtained. It is likely to be an RCB star based  on its lightcurve data.

The star, 6.6696.60, joins HV 12842 and W Men as members of the warm (6000-7000 K) RCB stars showing only weak molecular bands. 
The others are similar to HV 5637
which is typical of the cooler (5000 K) RCB stars which have much stronger molecular bands. See Figure 3. 
A low dispersion spectrum of 16.5641.22 (HV 2379) was previously obtained and is also very similar to HV 5637 
(Bessell \& Wood 1983).
The spectra were
examined for evidence of hydrogen by looking at the Balmer lines and the G-band of CH at 4300 \AA. The presence of 
$^{13}$C was searched for in the isotopic bands of C$_2$ and CN. 
In particular, the Swan bands, 
$^{12}$C$^{13}$C and $^{13}$C$^{13}$C near 4700 A, other C$_2$ bands in the 6000-6200 \AA\ region, and the $^{13}$CN 
band near 6250 A were examined.
The results are summarized in Table 3. 
 
Several other stars show irregular, fairly deep (2-3 mag) declines but fade much more slowly than typical 
RCB stars. In their lightcurve behavior, these stars resemble the unusual Galactic RCB star, DY Per (Alksnis
1994). 
This star has very deep  ($>$4 mag) declines at irregular intervals like an RCB star but the declines are very slow.
The DY Per declines appear much more symmetrical than the prototypical RCB decline which features a much faster drop
than rise.  Further spectroscopic analysis has shown that DY Per is very cool, T$_{eff}\sim$3500 K (Keenan \&
Barnbaum 1997). This is 
significantly cooler than the coolest known Galactic RCB stars, S Aps, WX CrA, and U Aqr which have estimated 
 T$_{eff}\sim$5000 K (Lawson et al. 1990).  Keenan \& Barnbaum (1997) suggest that DY Per may be hydrogen deficient.
The G-band is fairly weak. The evidence for the abundance of isotopic carbon is mixed.  The isotopic Swan bands, 
$^{12}$C$^{13}$C and $^{13}$C$^{13}$C near 4700 A are clearly seen but the $^{13}$CN band near 6250 A is not. 
The LMC DY Per stars share these characteristics. Four stars, 2.5871.1759, 10.3800.35, 15.10675.10, and 78.6460.7, show 
significant but slow declines of at least 2 magnitudes. See Figure 1d.  They also closely resemble DY Per spectroscopically.
See Figure 3c. The DY Per spectrum is from Barnbaum, Stone \& Keenan (1996).
Table 3 itemizes the major differences between the RCB and DY Per stars. 
The RCB declines are deeper and they fade faster as shown in the dm/dt column of the table. The DY Per stars show evidence for
significant amounts of  $^{13}$C and they are cooler than the RCB stars. 
The molecular bands in the cool RCB stars and the DY Per stars are of comparable strength 
but the DY Per spectra are intrinsically much redder.
The spectra of the DY Per stars resemble R-type carbon stars (Barnbaum et al. 1996). Many carbon 
stars also show irregular small ($\le$1 mag) declines. 
In addition, other stars produce dust such as carbon Miras and the V Hya stars. 
But these stars also show large regular variations not seen in either the
RCB or DY Per stars (e.g., Feast et al. 1984; Lloyd Evans 1997). 
Until more extensive observations and analysis are done, 
it is not clear whether
the DY Per stars are related to either the RCB stars or the carbon stars. 

\section{Individual Stars}

Using SIMBAD, each star was checked for previous identifications. 
These are included in Table 1. Previous observations from the literature for the sample stars are listed below.

\subsection{HV 12842}
This is the only star in the sample that lies in an area of the LMC not covered by MACHO.
It was first listed as an RCB star ($m_{pg}$ = 14.15-17.92 mag ) by
Payne-Gaposchkin (1971).
Several declines have been noted (Morgan, Nandy, \& Rao 1986; Lawson et al. 1990; Lawson, Cottrell, 
\& Pollard 1991). 
Coordinates are from the HST Guide Star Catalog.  HV 12842 is possibly an
IRAS source.  In the Faint Source Catalog, F05447-6425, lies within 
7\arcsec~of HV 12842.  At 12\micron, it is 0.09 $\pm$ 0.01 Jy. This is within a 
factor of 2 of the expected IRAS brightness of a Galactic RCB star seen at the 
distance of the LMC. 
This star had a median $m_{pg}$=15.2 mag and $\Delta$m = 2.5 mag in the
early 1900's
(Hodge \& Wright 1967 and references therein).

\subsection{20.5036.12 (HV 5637)}
This star was first listed as an RCB star by Hodge \&  Wright (1969). They found that 
HV 5637 (V$_{max}$= 14.99 mag , B$_{max}$ = 16.19 mag) had one decline of $\Delta$ B $\ge$ 2.36 mag around JD 2425000.
This star had a median $m_{pg}$=16.5 mag and $\Delta$m = 2.0 mag in the
early 1900's
(Hodge \& Wright 1967 and references therein).
It is listed as an RCB star ($m_{pg}$ = 16.38-18.20 mag) by
Payne-Gaposchkin (1971).
Butler (1978) noted small variations ($\Delta$V = 0.2 mag).
No further declines were noted during the seven years of MACHO coverage. See Figure 1. 

\subsection{21.7407.7 (W Men)}
This star was first identified as an RCB star by Luyten (1927).
Several declines have been noted (Milone 1975; Glass 1988; Lawson et al. 1990).
This star is HV 966 which had a median $m_{pg}$=14.4 mag and $\Delta$m = 1.2 mag in the
early 1900's
(Hodge \& Wright 1967 and references therein).
It is listed as an RCB star ($m_{pg}$ = 13.37-17.57 mag) by
Payne-Gaposchkin (1971).

\subsection{11.8632.2507 (HV 2671)}
This star had a median $m_{pg}$=16.4 mag and $\Delta$m = 1.8 mag in the
early 1900's
(Hodge \& Wright 1967 and references therein).
Kurochkin (1992) reports a maximum brightness of B= 15.5 mag and one deep decline with B $<$19 mag (JD 2439849.9 and 
2439852.75).

\subsection{6.6696.60 (HV 942)}
This star had a median $m_{pg}$=15.8 mag and $\Delta$m = 2.0 mag in the
early 1900's (Hodge \& Wright 1967 and references therein).
It is listed as an irregular variable of large range ($m_{pg}$ = 14.44-17.74 mag) by
Payne-Gaposchkin (1971).
This star is also possibly detected by IRAS (Schwering 1989). As with HV 12842, 
the IRAS brightness is consistent with an RCB at the distance of the LMC.

\subsection{16.5641.22 (HV 2379)}
This star had a median $m_{pg}$=16.5 mag and $\Delta$m = 1.8 mag in the
early 1900's
(Hodge \& Wright 1967 and references therein).
Wright \& Hodge (1971) report that in 1958, HV 2379 was seen at B=17.6 and then was below the plate limit for 95 days.  They
also summarize several hundred Harvard plates taken between 1896 and 1949 where the star varied between 16.2 and fainter than 
18.6 at B.  
It is listed as a long period  variable ($m_{pg}$ = 15.89-18.45 mag) by
Payne-Gaposchkin (1971).
Feast et al. (1984) suggest that HV 2379 is related to the carbon Mira, R For, but the MACHO lightcurve, shows a typical 
irregular RCB-star behavior with no sign of a Mira-type pulsation. 
This star is also possibly detected by IRAS (Trams et al. 1999). As with HV 12842, 
the IRAS brightness is consistent with an RCB at the distance of the LMC.
HV 2379 was also observed with ISO (van Loon 1999).

\subsection{18.3325.148 (HV 12524)}
This star had a median $m_{pg}$=15.9 mag and $\Delta$m = 0.6 mag in the
early 1900's
(Hodge \& Wright 1967 and references therein).
It is listed as an irregular variable of large range ($m_{pg}$ = 15.27-17.12mag ) by
Payne-Gaposchkin (1971).

\subsection{80.6956.207}
This star is also SHV 0523154-690100 ($<m_I>$ = 15.36 mag, $\Delta m_I$ = 1.16 mag) 
(Hughes 1989).

\subsection{80.7559.28}
This star is also SHV 0526537-690959 ($<m_I>$ = 14.74 mag, $\Delta m_I$ = 1.24 mag) 
(Hughes 1989).

\subsection{15.10675.10}
This star is also SHV 0546548-710843 ($<m_I>$ = 13.90 mag, $\Delta m_I$ = 1.14 mag) 
(Hughes 1989).

\section{Near-IR Colors}
Most of the sample has been observed one or more times in the near-IR. We have plotted the J-H vs. H-K 
colors in Figure 4. 
The typical RCB star colors evolve as dust forms and then disperses.  Feast (1997) shows the behavior of a large sample of 
Galactic RCB stars. 
The colors evolve from those typical of the RCB star photosphere toward those of a dust shell of 
$\sim$900 K. The colors for a combination of a 5500 K star and a 900 K shell are plotted in Figure 4.  The LMC RCB 
star colors are
consistent with the behavior of the Galactic RCB stars. 
With the exception of W Men, HV 2379, and HV 12842 (Bessell \& Wood 1983; Glass, Lawson, \& Laney 1994), each of the LMC stars 
has been observed only once or 
twice
at a random point in its lightcurve.  So, when plotted together, the ensemble of stars includes observations at maximum 
light and in declines. Together these observations map out an RCB star color evolution similar to that seen 
for the Galactic stars (Feast 1997).  Since there are a range of photospheric and shell temperatures,
Figure 4 shows more scatter than the plot of individual stars as shown by Feast (1997). Another interesting feature of
Figure 4 is that the DY Per stars are well separated from the RCB stars and show colors typical of carbon stars 
(Westerlund et al. 1991). The two DY Per stars plotted were observed at maximum light.

\section{Stellar Pulsations}
Most or possibly all of the RCB stars are pulsators (Lawson et al. 1990). As listed in Table 3 and seen in Figure 1, most of
the LMC RCB stars pulsate as well.  
We inspected the lightcurves for ``dormant" sections - places where
the lightcurve was within a magnitude or two of maximum brightness
and where, if it was changing, it was increasing slowly with
no obvious dust dropouts.
The long-term trend in
the dormant sections was subtracted, leaving just the shorter timescale
oscillations.
The baseline-subtracted, dormant sections of the V-band
lightcurves was run through the Discrete Fourier Transform Fortran code
of Roberts, Lehar, \& Dreher (1987).
The default frequency spacing was used, with a maximum period
of one per day.
In cases where there was an obvious signal, the
period corresponding to the frequency with the highest power is reported.
In most instances, the frequency spacing was about 5E-4 per day. Among the RCB stars, we were able to extract 
periods for 
18.3325.148 (83.8 d), 11.8632.2507 (60.0 d),
12.10803.56 (50.5 d), 21.7407.7 (240 d, questionable significance), and 79.5743.15 (53.3 d). For the DY Per stars, 
we extracted periods for
78.6460.7 (208 d), 
2.5871.1759 (138 d), 10.3800.35 (206 d),and 15.10675.10 (116 d).  
Theses two classes differ also in their typical pulsation period. The RCB stars in the LMC like their Galactic counterparts have
periods between 50 and 84 days while the periods of the DY Per stars lie between 100 and 210 days.
These differences are at least qualitatively in agreement with the results of theoretical models of 
pulsations in RCB stars (Weiss 1987). Theoretical periods for `Case 1' of Weiss
for M=0.825 M$_{\sun}$ show $\sim$40 d for T$_{eff} \sim$7000 K, $\sim$100 d for T$_{eff} \sim$5000 K, and 
$\sim$300 d for T$_{eff} \sim$4000 K.  Figure 32 of Lawson et al. (1990) shows the Galactic RCB stars plotted with Weiss
`Case 1' for comparison.

In a color-magnitude diagram (CMD), most of the of the RCB stars lie within the instability strip defined by
the MACHO-discovered BL Her, W Virs, and RV Tauri stars 
extrapolated to higher luminosities (Alcock et al. 2000). 
These LMC variables are
believed to be low-mass Population II stars.
In the CMD, the one hot RCB (11.8632.2507) lies blueward of the blue edge 
of the instability strip as defined by the bluest BL Her and W Vir stars.
The DY Per stars lie in the same region of the CMD as carbon-rich
red variables, and are brighter/redder than
the other observed sequences of red variables.
With the exception of 11.8632.2507, the RCB and DY Per stars
form an 
extension of the W Vir and BL Her star W-logP relation, where W = V - 2.0*(V-R), to higher luminosities 
and longer periods, although there may be a ``roll over" at long periods.

\section{Absolute Luminosity of RCB Stars} 
The Galactic
foreground reddening varies significantly across the face of the LMC ranging from E(B$-$V)$_{Gal}$=0.00 
to 0.17 mag (e.g., Schwering \& Israel 1991; Oestreicher, Gochermann, \& Schmidt-Kaler 1995).
Schwering \& Israel (1991) estimate that most of the LMC bar, where the MACHO fields are centered,  
has a Galactic foreground reddening of 
E(B-V)$_{Gal}$=0.06-0.08. The dust inside the LMC is patchy also but a good estimate of the reddening due to dust 
inside the LMC foreground to the RCB stars is E(B-V)$_{LMC}\sim$0.1 (Oestreicher \& Schmidt-Kaler 1996).
There will also be a small amount (E(B-V)$\sim$0.1 mag) of circumstellar reddening around each RCB star 
even at maximum light. We will ignore that component. The total reddening due to Galactic foreground and LMC intrinsic dust
is E(B-V)$\sim$0.17 mag or A$_V\sim$0.5 mag. 

The RCB and DY Per stars are plotted in Figure 5 in a V vs. V-R CMD. The values of V and V-R 
plotted are those measured for each star at maximum light. A reddening vector is also plotted. In addition, a line
is plotted representing the change in V and V-R with temperature assuming L$_V\propto~T^4$ with stellar radius held
fixed. Temperatures and colors are assumed to be those for normal supergiants (Cox 2000).
 On the basis of the three pre-MACHO RCB stars, Feast (1979) noted that the cool RCB star, HV 5637
  is significantly fainter than HV 12842 and W Men. He suggested that there might be a relationship between 
  absolute magnitude and effective temperature for the RCB stars.
The new observations reported here reinforce this suggestion.
The intrinsically brightest RCB stars at V are the warm ($\sim$7000 K) stars and the
faintest are the cool ($\sim$5000 K) stars.
There is a much wider range of absolute 
luminosity in the RCB stars than given by the canonical, $M_V$ = -4 to -5 mag.
The foreground reddening of individual stars is somewhat uncertain but the brightest RCB stars have $M_V$ $\sim$-5 mag and the
faintest are about $M_V$ $\sim$-2.5 mag.  The DY Per stars are cooler and fainter still, with a maximum absolute brightness of
$M_V$ $\sim$-2.5 mag.

It seems likely that 6.6696.60 is fairly heavily reddened. From Figure 3a, it can be seen that the spectrum of 
6.6696.60 closely resembles W Men and HV 12842. Therefore, it is likely that the effective temperatures and colors 
are similar for these three stars. For 6.6696.60, V$_{max}$= 15.0 and (V-R)$_{max}$=0.6 compared to V$_{max}$= 13.8 
and (V-R)$_{max}$=0.2 for the other two stars.  The simplest explanation is that some combination of circumstellar 
and interstellar dust in front of 6.6696.60 is responsible. The high value of reddening implied, E(B-V)$\sim$0.8 is 
unlikely to be primarily interstellar reddening. More likely, a large portion of this reddening is circumstellar. From
the lightcurve of 6.6696.60, seen in Figure 1b, it is quite possible that the star was never at 
maximum light during the
the seven years of the MACHO data. This is supported by earlier photometry of this star, previously 
discovered as HV 942,
which found the maximum brightness to be m$_{pg}$=14.4 (Payne-Gaposchkin 1971). This corresponds to B$\sim$14.3 mag. So 
assuming the same B-V color as W Men, then 6.6696.60 would have V$\sim$13.85 mag which is the same as W Men and HV 12842 at 
maximum 
light. Similarly, 10.3800.35 maybe more reddened than the other DY per stars.

\section{The Population of RCB stars}
The MACHO LMC sample of RCB stars allows us to attempt something not possible in the Milky Way, which is to estimate
the total population of RCB stars in a galaxy.
Figure 6 shows an image of the LMC with the locations of the RCB and DY Per stars plotted. 
As mentioned in the introduction there are two suggested evolutionary paths leading to the RCB stars, 
the Double Degenerate and the Final Helium Shell Flash (Iben et al. 1996a). 
Both these suggestions imply that the RCB stars are an old population.
The distribution of these stars on the sky and their radial velocities 
give clues to their origin. The space distribution and radial velocities of the Milky Way RCB
stars are similar to
those of distant planetary nebulae implying that these stars may be a bulge population (Drilling 
1986). 
However, the scale height is 400 pc for the RCB stars assuming M$_{Bol}$=-5 (Iben 
and Tutukov 1985). So the RCB 
stars may be more like old disk/Population I stars.
Either of the evolutionary scenarios predicts significantly more than the $\sim$30 RCB stars which are known in 
the Milky Way. For instance, Webbink
(1984) estimates a population of $\sim$1000 RCB stars making reasonable assumptions for the Double Degenerate scenario. 
However, it should be pointed out that Sch\"{o}nberner (1986) suggests that the estimated lifetimes for both scenarios 
are too
short to account for the number of RCB stars. 
Iben, Tutukov, \& Yungleson (1996b) estimate that final flashes could produce 30-2000 RCB stars at any given time 
depending on the core mass. They estimate that the Double Degenerate or binary merger scenario could produce $\sim$300 RCB 
stars.

Most of the Galactic RCB stars were discovered early in the century on the Harvard Observatory plate survey. 
The detection limit was about B= 11.8 mag on these blue sensitive plates. Stars near or below this limit, reddened stars and 
intrinsically red stars would have been missed (Lawson \& Cottrell 1990b; Lawson et al. 1990).  
The results for the LMC RCB stars imply that many Galactic RCB stars are 
significantly fainter than previously believed,
making them even less likely to have been detected. 

Using evolutionary models of the pulsation periods of 
RCB stars, one can estimate the crossing time
as a function of temperature (Lawson et al. 1990 and references therein). The model results imply relative 
populations of 
RCB stars with T$_{eff}$ =  5000, 6000 and 7000 K of 30:5:1. Most of the known RCB stars
in the Galaxy fall in the warmest subgroup. The observed population ratio is 1:1:4. Lawson et al.
(1990) suggest that the apparent lack of cool RCB stars is a selection effect. 
Lawson \& Cottrell (1990b) estimate that if all
RCB stars have M$_V \sim$-5, then the real number would be 200-300 or even larger if the number of cool stars equaled the 
number of warm stars.
The results of 
this study for the LMC imply that the ratio is 7:2:1 which is consistent with the theoretical ratio given above. 
If this is the intrinsic ratio in the Galaxy also, then there are 
$\sim10^3$ RCB stars in the entire Milky Way, most yet to be discovered. 

We can estimate the number of RCB stars in the LMC from the results of the MACHO sample.
The MACHO lightcurves for most stars have extremely good coverage for $\sim$2500 days spanning the years 1993 through 1999. 
This is long enough to catch most
but not all RCB stars in decline. 
RCB stars are true irregular variables.
The historical lightcurve of R CrB itself, which now stretches 
back over two hundred years, shows that it has gone up to 10 years with no decline and at other times has had several 
declines in one year (Mattei, Waagen, \& Foster 1991).
The characteristic time between declines is 1000-2000 days (Clayton, Whitney, \&
Mattei 1993).
So any search over a short time period will detect only a fraction of the RCB stars.
HV 5637 is an LMC RCB star but it is relatively inactive (Hodge \& Wright 1969).  It shows no
evidence for a decline during the 7 years of MACHO coverage. One of the Galactic RCB stars, XX Cam, shows similar
behavior. It has had only one recorded decline this century (Bidelman 1948; Yuin 1948). 
The AAVSO monitors 31 RCB stars including R CrB, V854 Cen and XX Cam. Over the same timespan covered by the MACHO 
observations, 23 of these stars or about three quarters of the observed sample had deep declines of 3 magnitudes or more. 
Another consideration 
is how long can an RCB star stay in a deep decline.  V854 Cen is a good example (Clayton 1996). 
It is 7th magnitude at maximum light yet 
it is not in the HD, CPD or SAO catalogs. Early in the 1900's, this star never rose above about magnitude 13.
This is a much rarer phenomenon. While several of the 31 RCB stars in the AAVSO sample were very active in the 1993-1999
time period, none spent the entire time in a deep decline. 
Therefore, using the Galactic RCB stars as a guide, we can expect to 
have detected about 75\% of the LMC RCB stars lying in the MACHO fields. 

Nine of the new RCB stars discovered with MACHO lie within the so-called top-22 fields which have the best temporal 
coverage 
(Alcock et al. 2000).  The three pre-MACHO LMC RCB stars and HV 2379 (16.5641.22) lie outside these fields.
The four DY Per stars
all lie in the top-22 fields. 
These 22 fields cover 10 square degrees centered on the LMC bar.  Kim et al. (1998) estimate the mass of the LMC 
within a 
radius of 4 kpc is 1.0 x $10^9$ $M_{\sun}$.  The inner 4 kpc of the LMC corresponds to 71.5 square degrees at a distance 
modulus of 18.4.  
Then, extrapolating from the top-22 MACHO fields,
we assume a homogeneous distribution of mass, and a 75\% detection efficiency.
This implies $\sim$85 RCB stars in the inner 4 kpc of the LMC.  We can then make an 
estimate of the expected number of RCB stars in the Galaxy assuming M$_{Galaxy} \sim$ 3.8 x 10$^{10}$ $M_{\sun}$. 
If RCB stars occur at the same frequency in the Galaxy as in the LMC and if the bar is not somehow special, then we would
expect there to be $\sim$3,200 RCB stars in the Milky Way. This number may be somewhat overestimated since the 
density of 
planetary 
nebulae and carbon stars is enhanced in the area of the LMC bar (e.g., Richer 1981; Morgan 1994). Similarly, 
we can estimate the expected number in the 
SMC. The top-6 MACHO fields in the SMC were searched in a similar manner to that described above for the LMC. These 6 
fields in
the SMC contain 2.2 x 10$^6$ stars as compared to 8.5 x 10$^6$ stars in the top-22 LMC fields 
(Alcock et al. 1997b). So extrapolating from the number
found in the LMC, we would expect to find $\sim$2 RCB stars in the SMC. None were found. 

\section{Summary}
The final results of our study of the LMC RCB stars in the top-22 MACHO fields are:\\
$\bullet$ Ten new RCB stars (eight reported in this paper) were discovered using MACHO photometry bringing 
the total in the LMC to thirteen.\\
$\bullet$ There is a much wider range of absolute luminosity in the RCB stars than previously thought.
There is a range of $M_V$ $\sim$-2.5 to -5 mag. The warm ($\sim$7000 K) stars are brighter at V than the
cool ($\sim$5000 K) stars. \\
$\bullet$ The relative populations of LMC RCB stars with T$_{eff}$ =  5000, 6000 and 7000 K is 7:2:1 close to the theoretical 
ratio but very different from the Galactic ratio of 1:1:4. The Galactic ratio may be severely biased due to selection effects.\\
$\bullet$ One hot ($\sim$20,000 K) RCB star was found. Three such stars are known in the Galaxy.\\
$\bullet$ If the population of RCB stars in the LMC is similar to that of the Galaxy then there may be $\sim$3,200 RCB stars
in the Galaxy.\\
$\bullet$ Four stars were found which resemble the unusual Galactic RCB star, DY Per.  These stars fade much more slowly than typical
RCB stars. 
These stars are also cooler and fainter, with a maximum absolute brightness of
$M_V$ $\sim$-2.5 mag.
It is not clear if these stars are related to the RCB stars.

\acknowledgments

  We are very grateful for the skilled support given our project by the technical staffs
at the Mount Stromlo and CTIO Observatories, and, in particular, we would like to
thank Simon Chan, Glen Thorpe, Susannah Sabine, and Michael McDonald for
their invaluable assistance in obtaining the data. Work performed at the
University of California Lawrence Livermore National Laboratory is supported by
the US Department of Energy under contract W7405-Eng-48. Work performed by
the Center for Particle Astrophysics personnel is supported in part by the Office of
Science and Technology Centers of the NSF under cooperative agreement
A-8809616. Work performed at MSSSO is supported by the Bilateral Science and
Technology Program of the Australian Department of Industry, Technology and
Regional Development. DM is supported by Fondecyt 1990440. 
We are grateful to the SAAO and CTIO for providing time for follow-up 
observations. 
We thank Peter Wood for suggesting that HV 2379 might be an RCB star and Cecilia Barnbaum for 
providing the
spectrum of DY Per. This work made use of the SIMBAD astronomical database provided by the
Centre de Donn\`{e}es astronomiques de Strasbourg.

\section{Appendix}
In this Appendix we provide the MACHO photometric data for each of the stars in the sample. The data are given 
in Tables A1 to
A16 and include the HJD, R Magnitude, V magnitude and the Observation ID which is the running exposure number. 
The magnitudes 
are calculated as described in Section 2.1.  In the paper edition, only a portion of Table A1 is given as an 
example. The 
remaining tables appear only in the electronic edition.

\begin{table*}
\scriptsize
\begin{center}
\caption{LMC RCB Stars }
\begin{tabular}{llccc}
$Name$ & MACHO Name& $\alpha(2000)$&
 $\delta(2000)$&
$Other ID$\\
\tableline
\tableline
&&&&\\
\underline{Previously known RCB Stars}&&&&\\
HV 12842&Outside MACHO Field   &05 45 02.92     &-64 24 23.2    &F05447-6425$^a$\\
HV 5637 (20.5036.12)&MACHO*05:11:31.4-67:55:52&05 11 31.402 & -67 55 52.34&SP 37-16$^b$\\
W Men  (21.7407.7) &MACHO*05:26:24.7-71:11:13& 05 26 24.761 &-71 11 13.32   &HV 966, R102\\
11.8632.2507 &MACHO*05:33:49.1-70:13:22$^c$&05 33 49.126& -70 13 22.49& hot RCB star, HV 2671\\
81.8394.1358 &MACHO*05:32:13.3-69:55:59$^c$&05 32 13.263  &   -69 55 59.23&\\
&&&&\\
\underline{Confirmed new RCB Stars}&&&&\\
6.6575.13 &MACHO*05:20:48.2-70:12:12&05 20 48.244  &   -70 12 12.51&\\
6.6696.60&MACHO*05:21:47.9-70:09:57&05 21 47.997    & -70 09 57.41&HV 942\\
12.10803.56&MACHO*05:46:46.2-70:38:12&   05 46 46.211   &  -70 38 12.79  &\\
16.5641.22&MACHO*05:14:46.2-67:55:48&05 14 46.177   &-67 55 48.07& HV 2379 \\
18.3325.148       &MACHO*05:01:00.2-69:03:43& 05 01 00.299  &   -69 03 43.53&HV 12524, SP 31-38$^b$\\
79.5743.15&MACHO*05:15:51.8-69:10:08&05 15 51.834   &  -69 10 08.99 &PMN J0515-6910$^d$\\
80.6956.207&MACHO*05:22:56.9-68:58:16&   05 22 56.944  &   -68 58 16.87 &SHV 0523154-690100$^e$\\
80.7559.28&MACHO*05:26:33.8-69:07:33&05 26 33.853  &   -69 07 33.21&SHV 0526537-690959$^e$\\
&&&&\\
\underline{DY Per type Stars}&&&&\\
2.5871.1759  &MACHO*05:16:51.8-68:45:17& 05 16 51.890  &   -68 45 17.81&\\
10.3800.35       &MACHO*05:03:44.7-69:38:12& 05 03 44.790  &   -69 38 12.73&\\
15.10675.10     &MACHO*05:46:13.0-71:07:40& 05 46 13.045  &   -71 07 40.62&SHV 0546548-710843$^e$, SP 55-23$^b$\\
78.6460.7          &MACHO*05:19:56.0-69:48:06& 05 19 56.050  &   -69 48 06.44&\\

\tableline
\end{tabular}
\tablenotetext {a}{IRAS faint Source Catalog}
\tablenotetext {b}{Carbon star (Sanduleak, \& Philip 1977)}
\tablenotetext {c}{Discovered previously by Alcock et al. (1996).
Finding chart is given there.}
\tablenotetext {d}{Parkes-MIT-NRAO Survey. Flux at 4850 MHz is $62 \pm 7$ mJy 
centered 5.3\arcsec~from the star (Wright et al. 1994).}
\tablenotetext {e}{Previously discovered as an LPV in the LMC (Hughes 1989).}

\end{center}
\end{table*}

\begin{table*}
\scriptsize
\begin{center}
\caption{New LMC RCB Photometry}
\begin{tabular}{llllllll}
Name&JD&V&R&J&H&K&Ref.\\
\tableline
\tableline
&&&&&\\
HV 12842&2451158&--&--&12.87&12.53&11.79&2\\
HV 5637 (20.5036.12)&2451113&14.8&14.1&12.84&12.57&12.18&2\\
W Men (21.7407.7)&2450906&14.2&14.4&13.15&12.92&12.22&2\\
10.3800.35&2451112&17.4&15.6&12.18&10.91&10.17&2\\
11.8632.2507 &2450448&16.0&15.9&15.29&14.61&12.93&1\\
 &2450892&16.0&15.8&15.07&14.13&12.83&2\\
12.10803.56&2450448&15.2&14.4&12.74&12.24&11.76&1\\
&2450894&15.3&14.5&12.74&12.19&11.67&2\\
 15.10675.10&2450894&16.2&15.1&12.22&11.08&10.50&2\\
16.5641.22&2451113&16.6&15.6&13.44&12.75&11.90&2\\
18.3325.148&2451112&14.6&13.9&12.71&12.37&12.08&2\\
79.5743.15&2450448&15.6&14.6&12.79&12.01&11.15&1\\
&2451113&19.1&18.2&15.43&13.96&12.42&2\\
80.6956.207&2450448&17.5&16.4&14.32&13.27&12.33&1\\
&2450906&$\sim$21&20.2&17.70&14.64&12.64&2\\
80.7559.28&2450448&16.4&15.2&13.03&12.17&11.35&1\\
&2450906&20.2&18.9&15.75&14.17&12.43&2\\
81.8394.1358&2450448&17.1&16.2&13.71&13.04&12.28&1\\
&2450906&18.8&18.0&15.02&13.92&12.62&2\\
\tableline
\end{tabular}
\tablenotetext {1}{This paper, CTIO}
\tablenotetext {2}{2MASS}
\end{center}
\end{table*}

\begin{table*}
\scriptsize
\begin{center}
\caption{Final Identification}
\begin{tabular}{lllllllllll}
$Name$& $V_{max}$& $R_{max}$&(V-R)& $\Delta$m&$\Delta$t&dm/dt&$^{13}$C&H/CH&Pulsations&Comment\\
\tableline
\tableline
&&&&&\\
\underline{RCB Stars}&&&&&&&&&&\\
HV 12842		&13.70	&13.50	&0.20	&$>$4.0	&53	&0.07&none&weak CH&yes&\\
HV 5637 (20.5036.12)	&14.75	&14.10	&0.65	&$>$2.4	&--	&--&none&weak CH&?&Only one decline known\\
W Men (21.7407.7)	&13.90	&13.67	&0.23	&2.70	&30	&0.09&none&none?&yes (240: d)&possible H$\beta$\\
6.6575.13	&15.25	&14.50	&0.75	&4.00	&50	&0.08&--&--&yes&no spectra at all\\
6.6696.60	&15.00	&14.40	&0.60	&5.40	&45	&0.12&none&none&yes&\\
11.8632.2507	&16.10	&16.00	&0.10	&3.00	&25	&0.12&--&--&yes (60.0 d)&Only Hot RCB in the LMC\\
12.10803.56	&15.10	&14.25	&0.85	&5.80	&75	&0.08&none&none&yes (50.5 d)&No red spectra\\
79.5743.15	&15.20	&14.30	&0.90	&4.60	&50	&0.09&none&weak CH&yes (53.3 d)&no red spectra\\
16.5641.22      &14.90  &14.10  &0.80   &6.50   &125    &0.06 &none?&none?&?&spectrum (Bessell \& Wood 1983)\\
18.3325.148	&14.50	&13.80	&0.70	&$>$1.7	&--	&--    &none&weak CH&yes (83.8 d)&No declines known\\
80.6956.207	&16.00	&15.00	&1.00	&5.40	&60	&0.09&yes&none&yes&similar to DY Per\\
80.7559.28	&15.80	&14.75	&1.05	&5.75	&35	&0.16&none&none&yes&no red spectra\\
81.8394.1358	&16.30	&15.50	&0.80	&3.50	&50	&0.06&none&none&?&no red spectra\\

&&&&&&&&&\\
\underline{DY Per Stars}&&&&&&&&&\\
2.5871.1759	&16.10	&14.85	&1.25	&2.10	&200	&0.01&yes&none&yes (138 d)&\\
10.3800.35	&17.40	&15.70	&1.70	&3.20	&275	&0.01&none&none&yes (206 d)&no flux at CH or 4700\AA\\
15.10675.10	&16.00	&14.80	&1.20	&2.10	&400	&0.005&yes&weak H$\beta$?&yes (116 d)&no flux at CH\\
78.6460.7	&16.00	&14.45	&1.55	&2.10	&300	&0.007&none&none&yes (208 d)&no flux at CH or 4700 A \\
\tableline
\end{tabular}
\end{center}
\end{table*}

\begin{deluxetable}{cccr}
%\small
\scriptsize
\tablenum{A1}
\tablecolumns{4}
\tablewidth{0pc}
\tablecaption{MACHO Project Photometry for 20.5036.12\label{tab-20.5036.12}} 
\tablehead{
\colhead{HJD - 2,400,000}&
\colhead{R}&
\colhead{V}&
\colhead{ObsID}
\\
&\colhead{\sl (mag)}&\colhead{\sl (mag)}&}
\startdata
{\tt48825.3106}&{\tt14.243}$\pm${\tt0.016}&{\tt14.859}$\pm${\tt0.017}&{\tt   336}\\
{\tt48833.2274}&{\tt14.189}$\pm${\tt0.015}&{\tt14.866}$\pm${\tt0.015}&{\tt   508}\\
{\tt48842.2419}&{\tt14.230}$\pm${\tt0.015}&{\tt14.939}$\pm${\tt0.015}&{\tt   765}\\
{\tt48843.2827}&{\tt14.249}$\pm${\tt0.015}&{\tt14.941}$\pm${\tt0.015}&{\tt   802}\\
{\tt48844.2223}&{\tt14.237}$\pm${\tt0.015}&{\tt14.919}$\pm${\tt0.015}&{\tt   837}\\
{\tt48852.2494}&{\tt14.226}$\pm${\tt0.015}&{\tt14.904}$\pm${\tt0.015}&{\tt   916}\\
{\tt48856.3140}&{\tt14.342}$\pm${\tt0.017}&{\tt14.997}$\pm${\tt0.016}&{\tt  1051}\\
{\tt48928.1216}&{\tt14.159}$\pm${\tt0.015}&{\tt14.845}$\pm${\tt0.015}&{\tt  1633}\\
{\tt48966.1570}&{\tt14.236}$\pm${\tt0.015}&{\tt14.922}$\pm${\tt0.015}&{\tt  2237}\\
{\tt48967.1506}&{\tt14.214}$\pm${\tt0.015}&{\tt14.906}$\pm${\tt0.015}&{\tt  2283}\\
{\tt49096.0556}&{\tt14.195}$\pm${\tt0.015}&{\tt14.893}$\pm${\tt0.015}&{\tt  5421}\\
{\tt49126.9777}&{\tt14.094}$\pm${\tt0.015}&{\tt14.769}$\pm${\tt0.015}&{\tt  7271}\\
{\tt49135.9786}&{\tt14.117}$\pm${\tt0.015}&{\tt14.799}$\pm${\tt0.015}&{\tt  7783}\\
{\tt49165.2826}&{\tt14.222}$\pm${\tt0.015}&{\tt14.899}$\pm${\tt0.015}&{\tt  9111}\\
{\tt49184.3319}&{\tt14.127}$\pm${\tt0.015}&{\tt14.801}$\pm${\tt0.015}&{\tt  9835}\\
{\tt49190.2491}&{\tt14.150}$\pm${\tt0.015}&{\tt14.814}$\pm${\tt0.015}&{\tt 10257}\\
{\tt49204.2702}&{\tt14.211}$\pm${\tt0.015}&{\tt14.965}$\pm${\tt0.015}&{\tt 10743}\\
{\tt49212.2431}&{\tt14.192}$\pm${\tt0.015}&{\tt14.910}$\pm${\tt0.015}&{\tt 11203}\\
{\tt49225.2646}&{\tt14.257}$\pm${\tt0.015}&{\tt15.109}$\pm${\tt0.015}&{\tt 11977}\\
{\tt49234.2266}&{\tt14.220}$\pm${\tt0.015}&{\tt14.893}$\pm${\tt0.015}&{\tt 12109}\\
{\tt49290.2091}&{\tt14.167}$\pm${\tt0.015}&{\tt14.843}$\pm${\tt0.015}&{\tt 13296}\\
{\tt49313.1389}&{\tt14.087}$\pm${\tt0.015}&{\tt14.756}$\pm${\tt0.015}&{\tt 13650}\\
{\tt49376.1739}&{\tt14.180}$\pm${\tt0.015}&{\tt14.875}$\pm${\tt0.015}&{\tt 14016}\\
{\tt49381.1787}&{\tt14.115}$\pm${\tt0.015}&{\tt14.800}$\pm${\tt0.015}&{\tt 14235}\\
{\tt49399.1561}&{\tt14.055}$\pm${\tt0.015}&{\tt14.737}$\pm${\tt0.015}&{\tt 14550}\\
{\tt49406.2353}&{\tt14.102}$\pm${\tt0.015}&{\tt14.801}$\pm${\tt0.015}&{\tt 14715}\\
{\tt49425.0567}&{\tt14.090}$\pm${\tt0.015}&{\tt14.775}$\pm${\tt0.015}&{\tt 15182}\\
{\tt49437.1389}&{\tt14.089}$\pm${\tt0.015}&{\tt14.780}$\pm${\tt0.015}&{\tt 15661}\\
{\tt49452.1000}&{\tt14.050}$\pm${\tt0.015}&{\tt14.735}$\pm${\tt0.015}&{\tt 16284}\\
{\tt49478.0438}&{\tt14.163}$\pm${\tt0.015}&{\tt14.841}$\pm${\tt0.015}&{\tt 17406}\\
{\tt49513.3355}&{\tt14.252}$\pm${\tt0.015}&{\tt14.936}$\pm${\tt0.015}&{\tt 19015}\\
{\tt49547.2278}&{\tt14.146}$\pm${\tt0.015}&{\tt14.810}$\pm${\tt0.015}&{\tt 20717}\\
{\tt49559.3172}&{\tt14.154}$\pm${\tt0.015}&{\tt14.824}$\pm${\tt0.015}&{\tt 21553}\\
{\tt49573.2564}&{\tt14.222}$\pm${\tt0.015}&{\tt14.898}$\pm${\tt0.015}&{\tt 22350}\\
{\tt49608.1654}&{\tt14.234}$\pm${\tt0.015}&{\tt14.916}$\pm${\tt0.015}&{\tt 24240}\\
{\tt49631.1699}&{\tt14.217}$\pm${\tt0.015}&{\tt14.966}$\pm${\tt0.015}&{\tt 25388}\\
{\tt49650.1537}&{\tt14.349}$\pm${\tt0.015}&{\tt15.113}$\pm${\tt0.015}&{\tt 26097}\\
{\tt49705.1947}&{\tt14.137}$\pm${\tt0.015}&{\tt14.818}$\pm${\tt0.015}&{\tt 27375}\\
{\tt49731.1965}&{\tt14.149}$\pm${\tt0.015}&{\tt14.845}$\pm${\tt0.015}&{\tt 27730}\\
{\tt49742.9844}&{\tt14.164}$\pm${\tt0.015}&{\tt14.870}$\pm${\tt0.015}&{\tt 27905}\\
{\tt49748.9535}&{\tt14.145}$\pm${\tt0.015}&{\tt14.851}$\pm${\tt0.015}&{\tt 28093}\\
{\tt49752.1860}&{\tt14.167}$\pm${\tt0.015}&{\tt14.851}$\pm${\tt0.015}&{\tt 28230}\\
{\tt49761.1006}&{\tt14.202}$\pm${\tt0.015}&{\tt14.904}$\pm${\tt0.015}&{\tt 28451}\\
{\tt49766.2242}&{\tt14.120}$\pm${\tt0.015}&{\tt14.812}$\pm${\tt0.016}&{\tt 28594}\\
{\tt49769.2219}&{\tt14.086}$\pm${\tt0.015}&{\tt14.778}$\pm${\tt0.015}&{\tt 28656}\\
\enddata
\end{deluxetable}

\begin{figure*}
\figurenum{1a}
\epsscale{0.75}
\plottwo{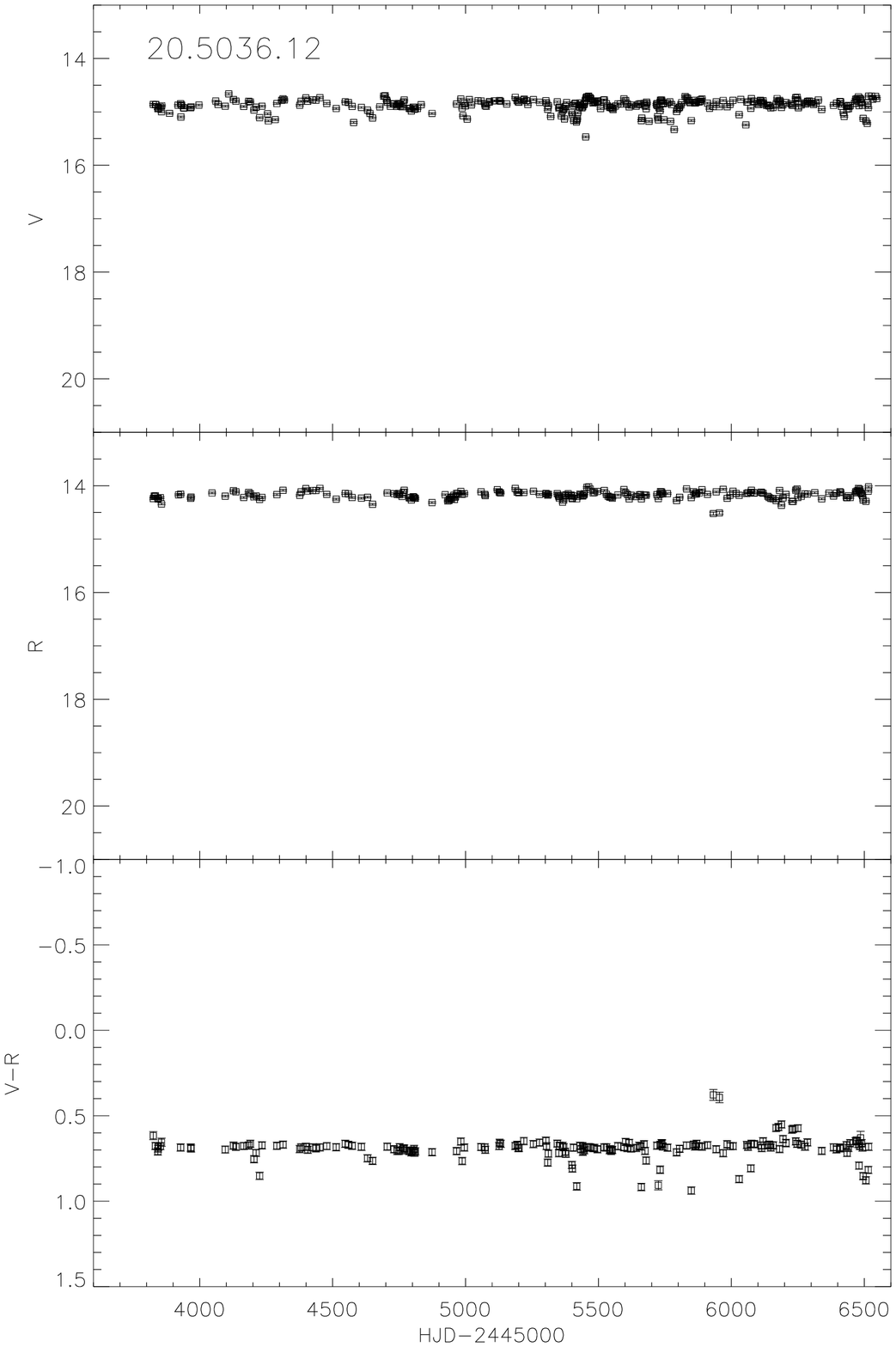}{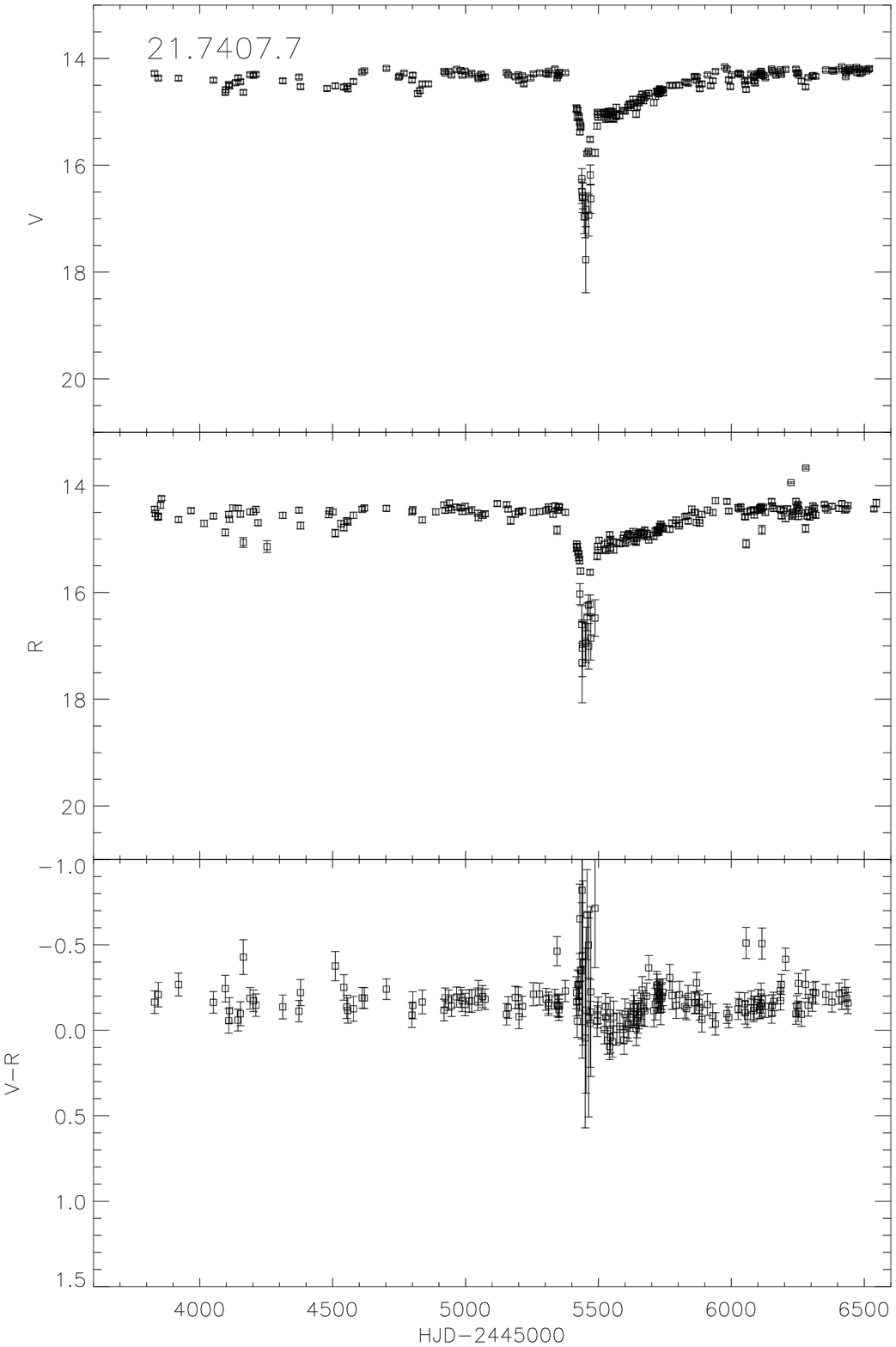}\\
\plottwo{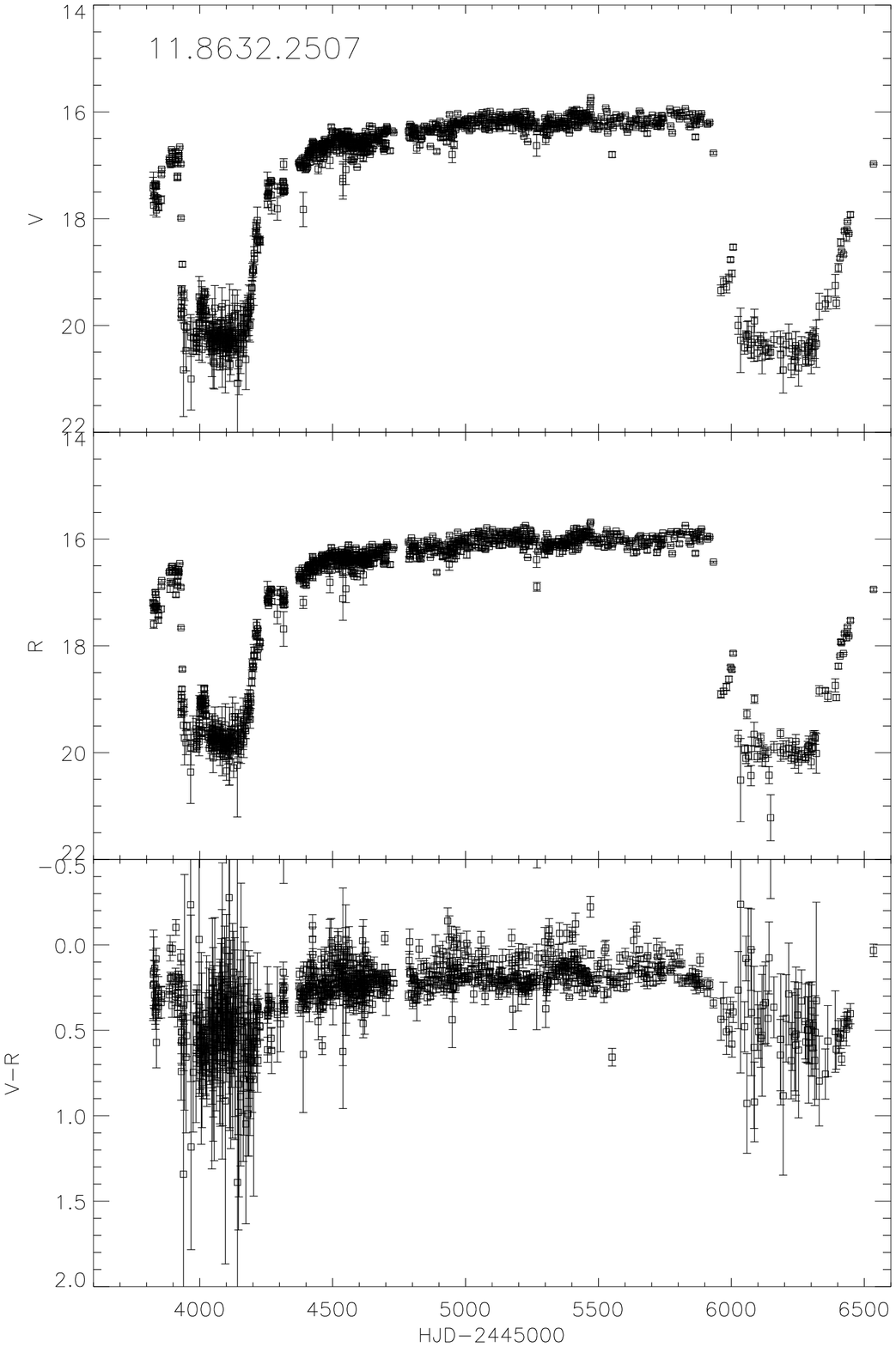}{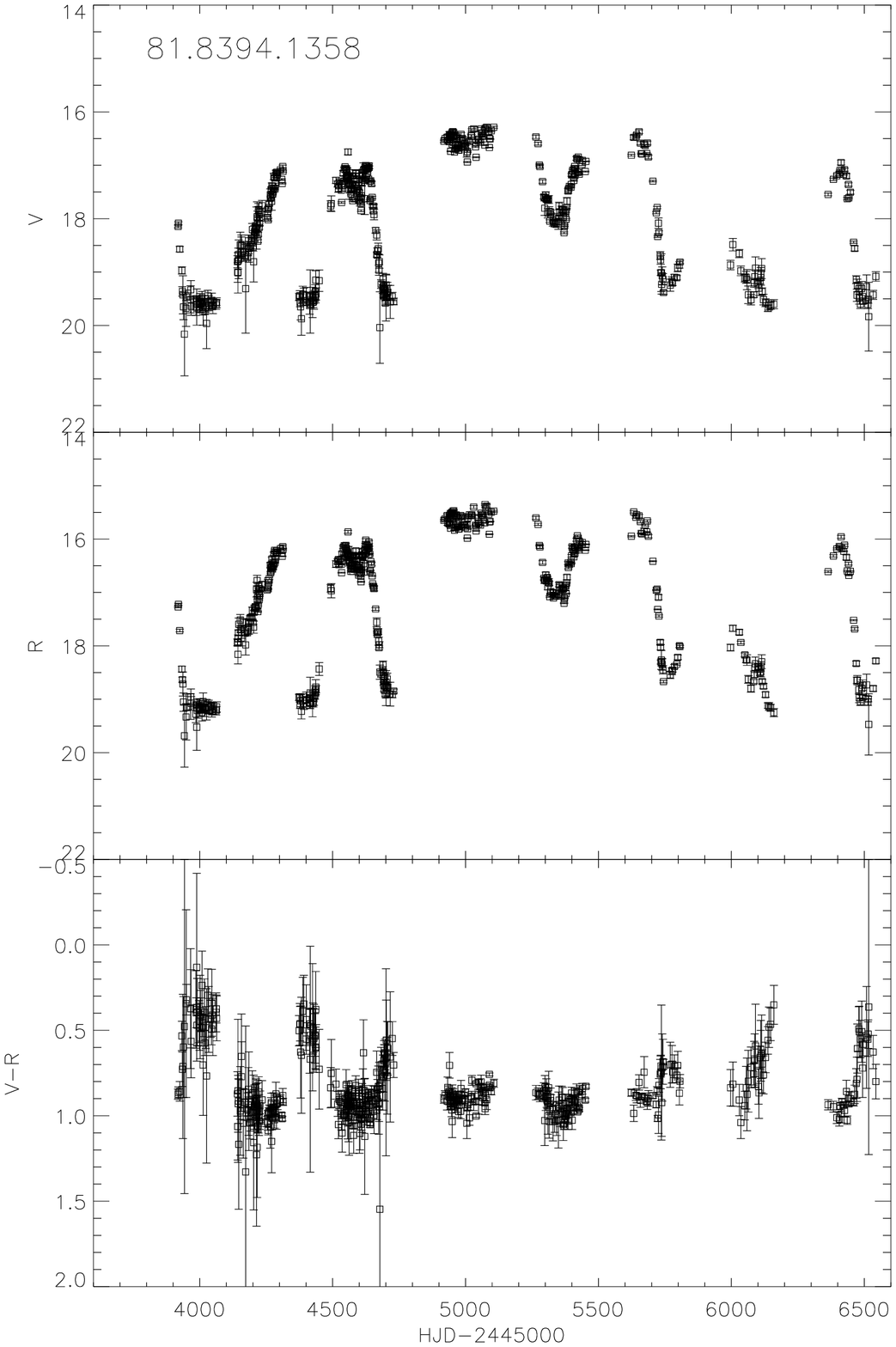}
\caption{MACHO lightcurves for confirmed RCB stars. }
%\epsscale{1.0}
\end{figure*}

\begin{figure*}
\figurenum{1b}
\epsscale{0.75}
\plottwo{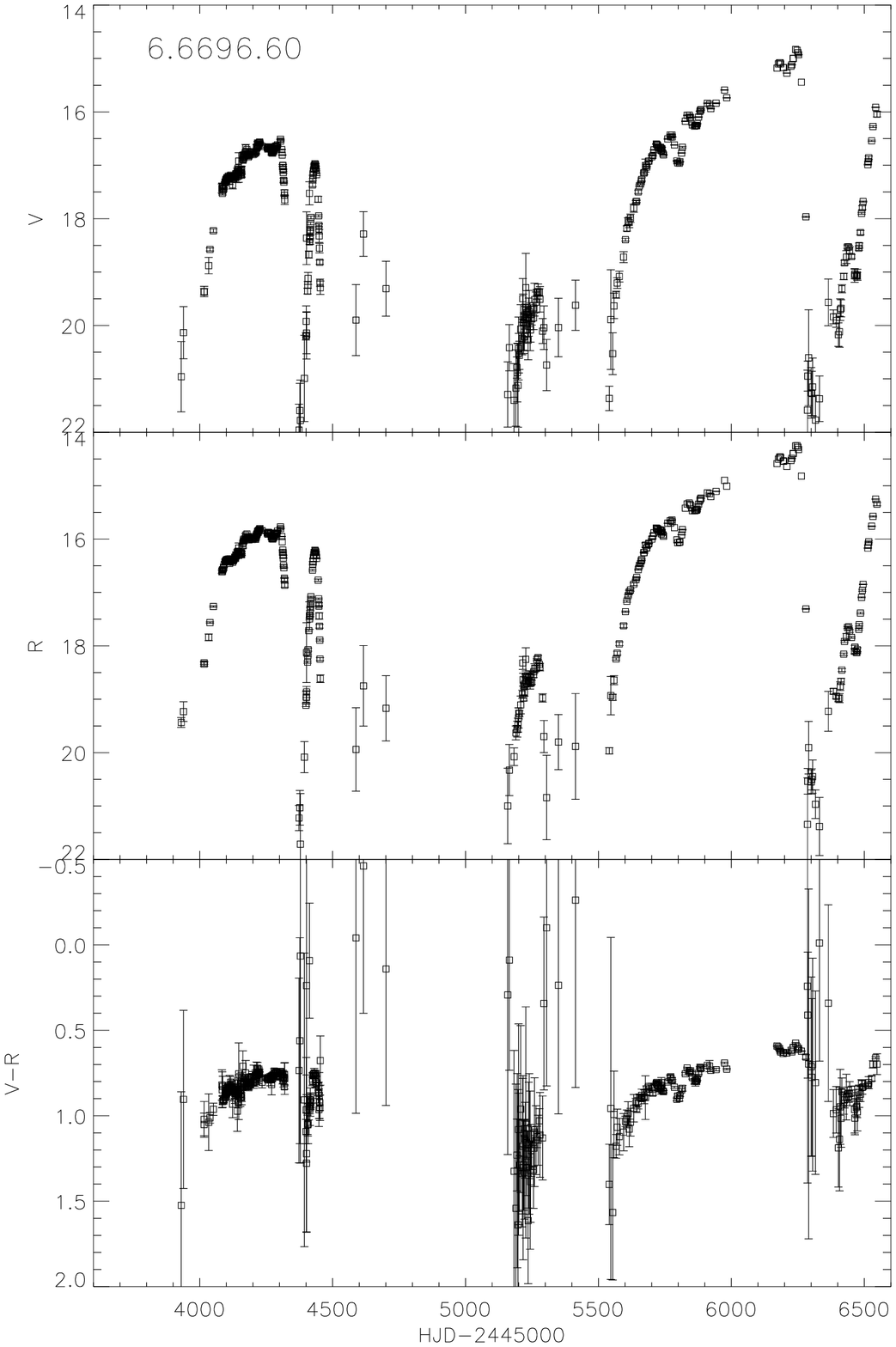}{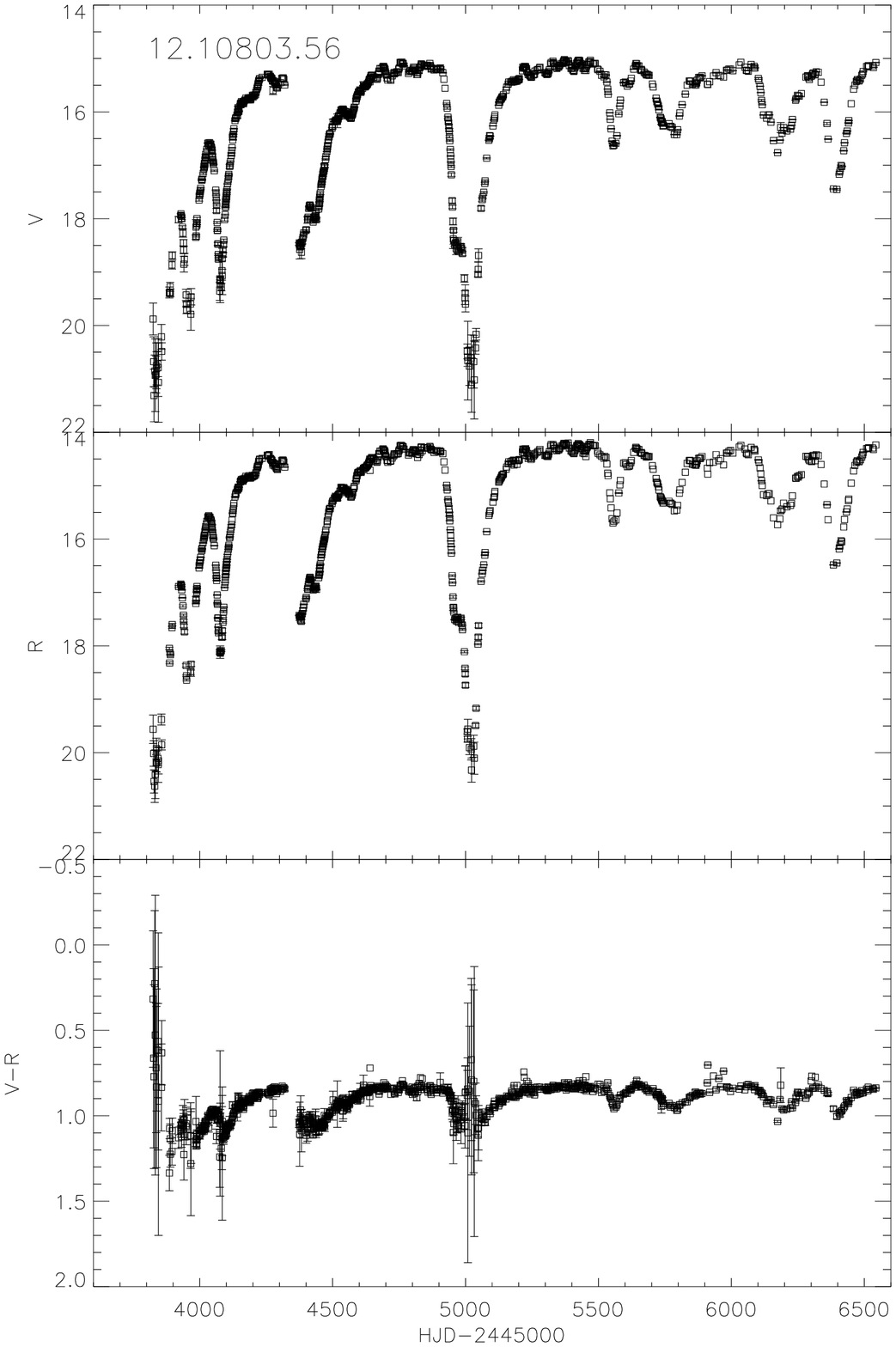}\\
\plottwo{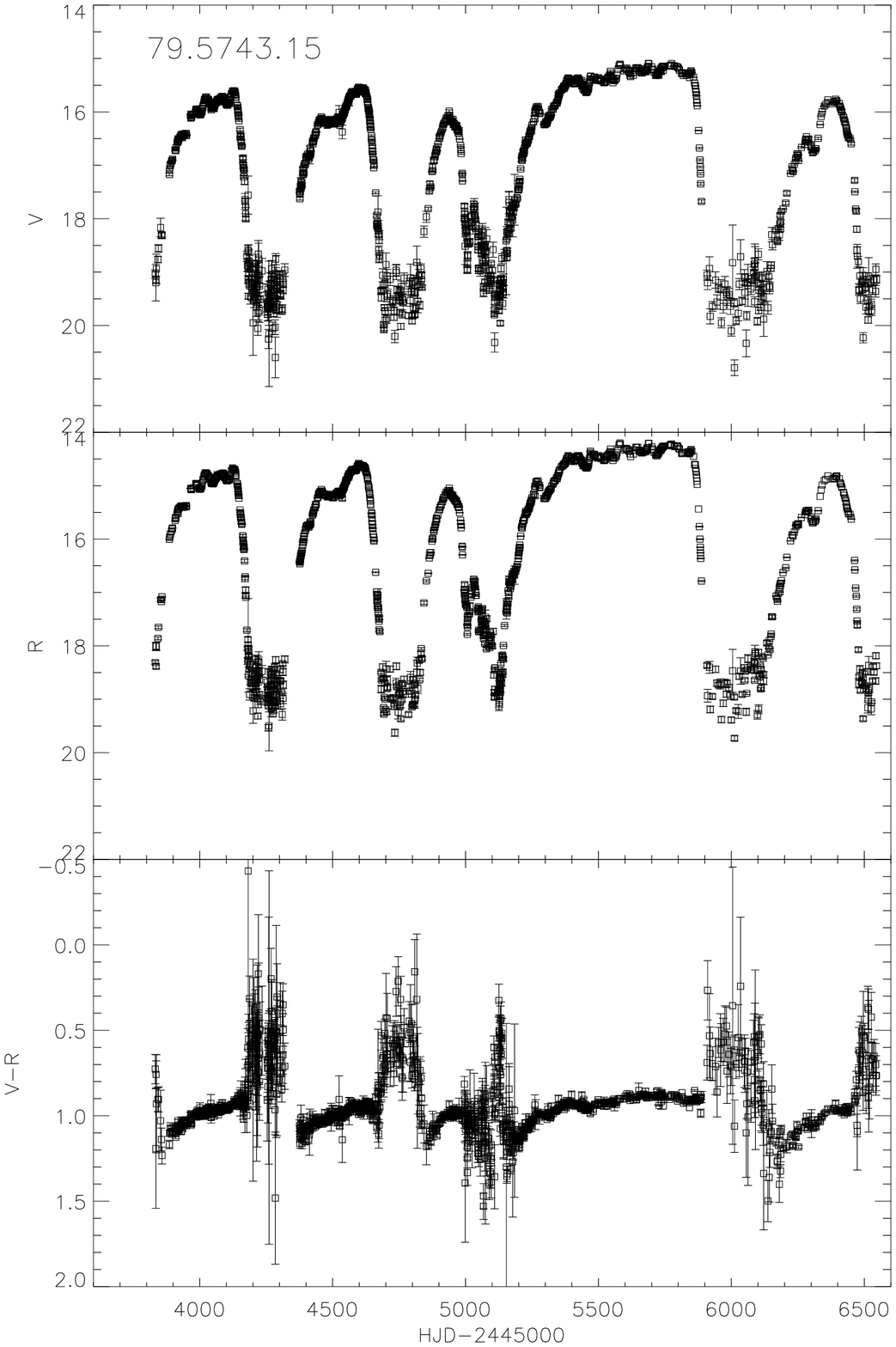}{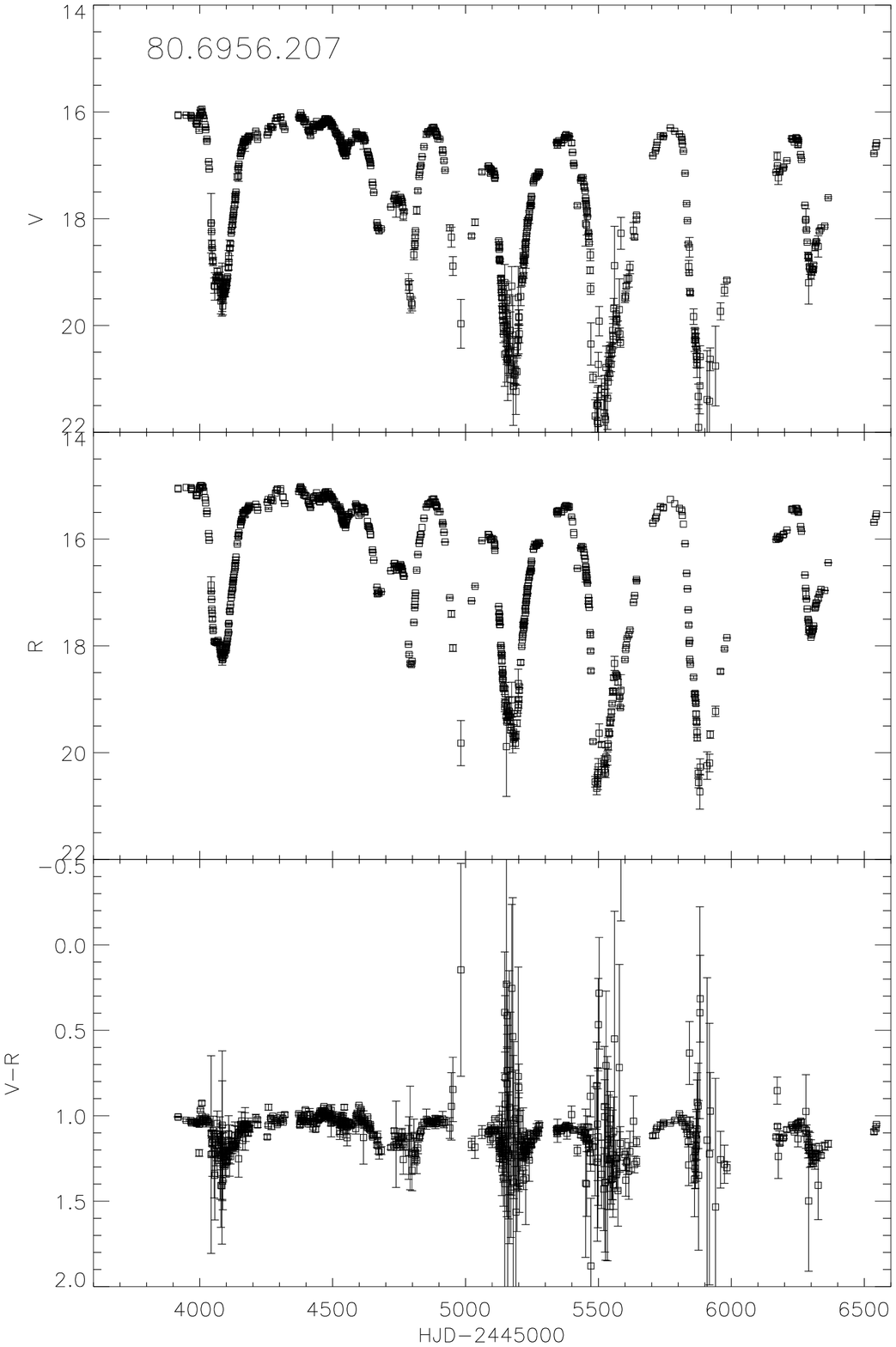}\\
\caption{MACHO lightcurves for confirmed RCB stars. }

\end{figure*}

\begin{figure*}
\figurenum{1c}
\epsscale{0.75}
\plottwo{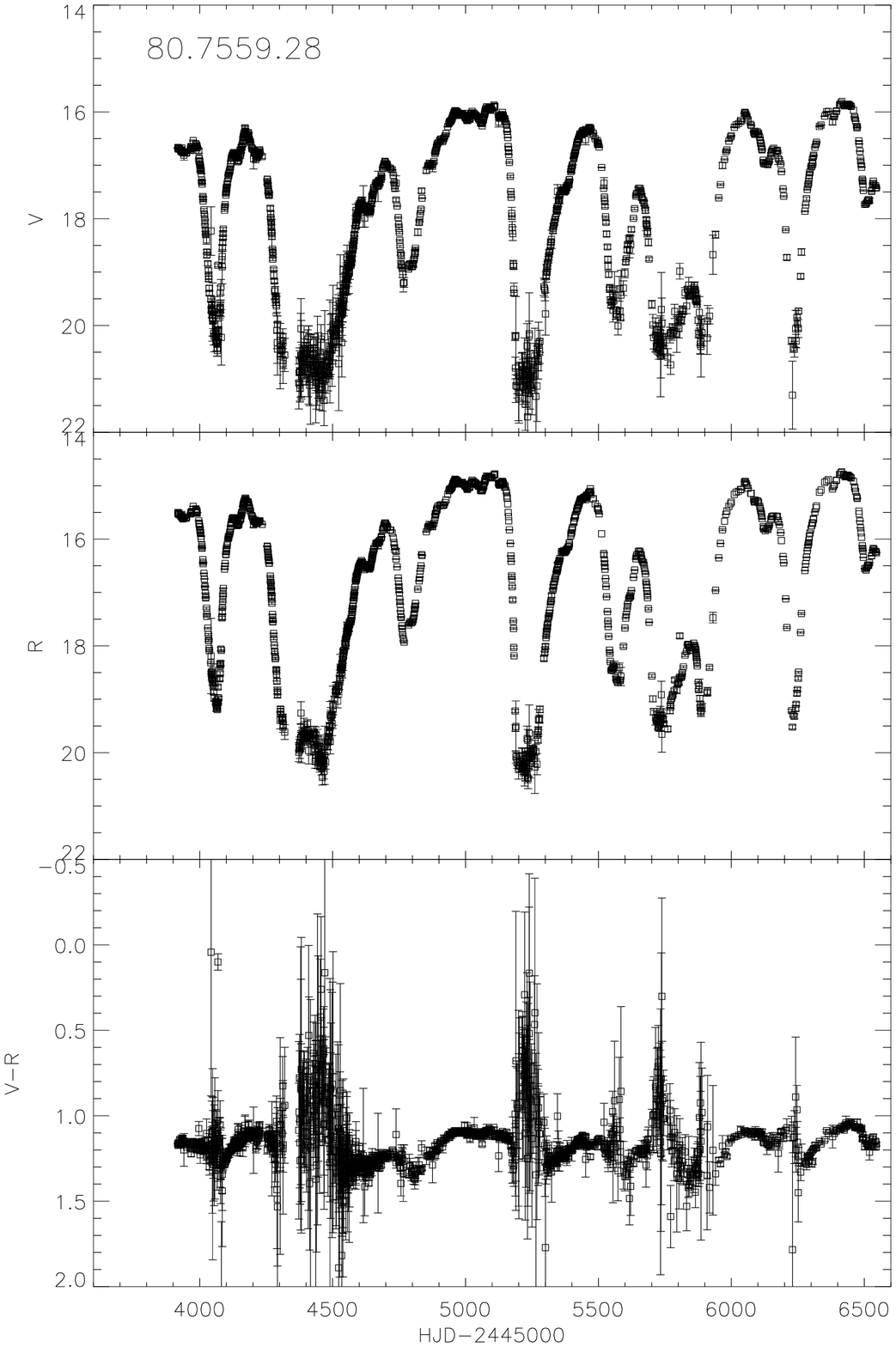}{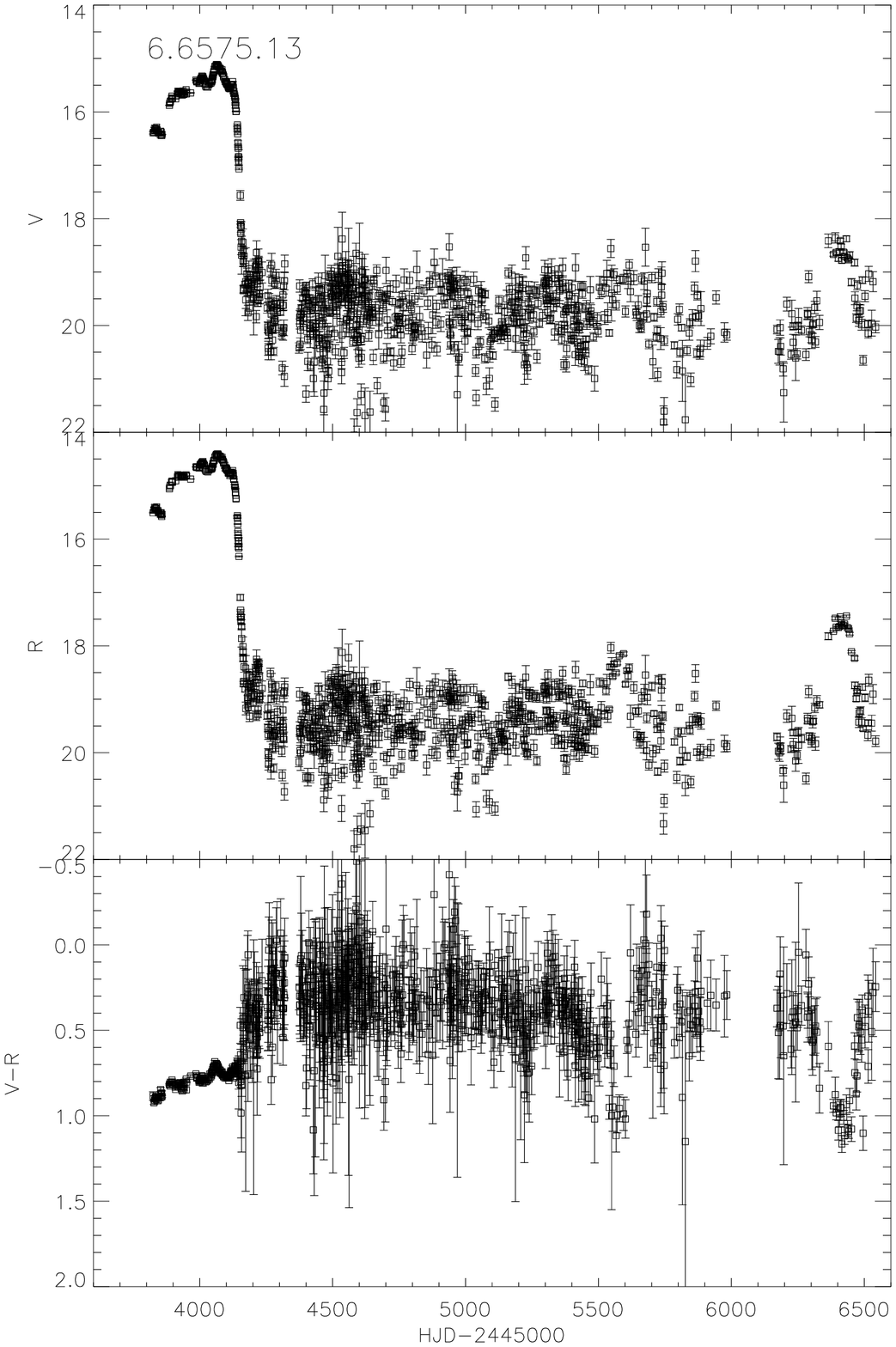}\\
\epsscale{0.75}
\plottwo{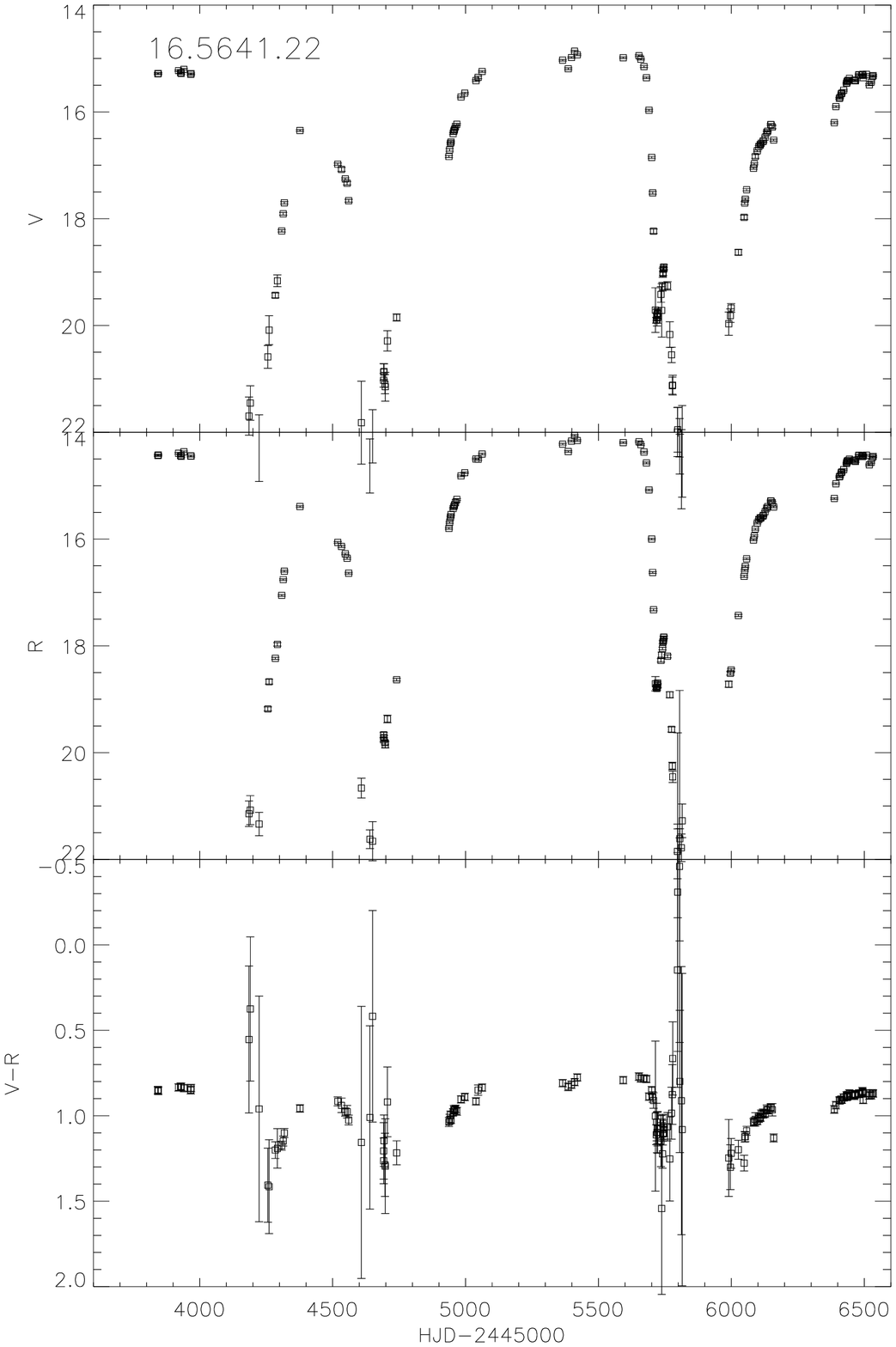}{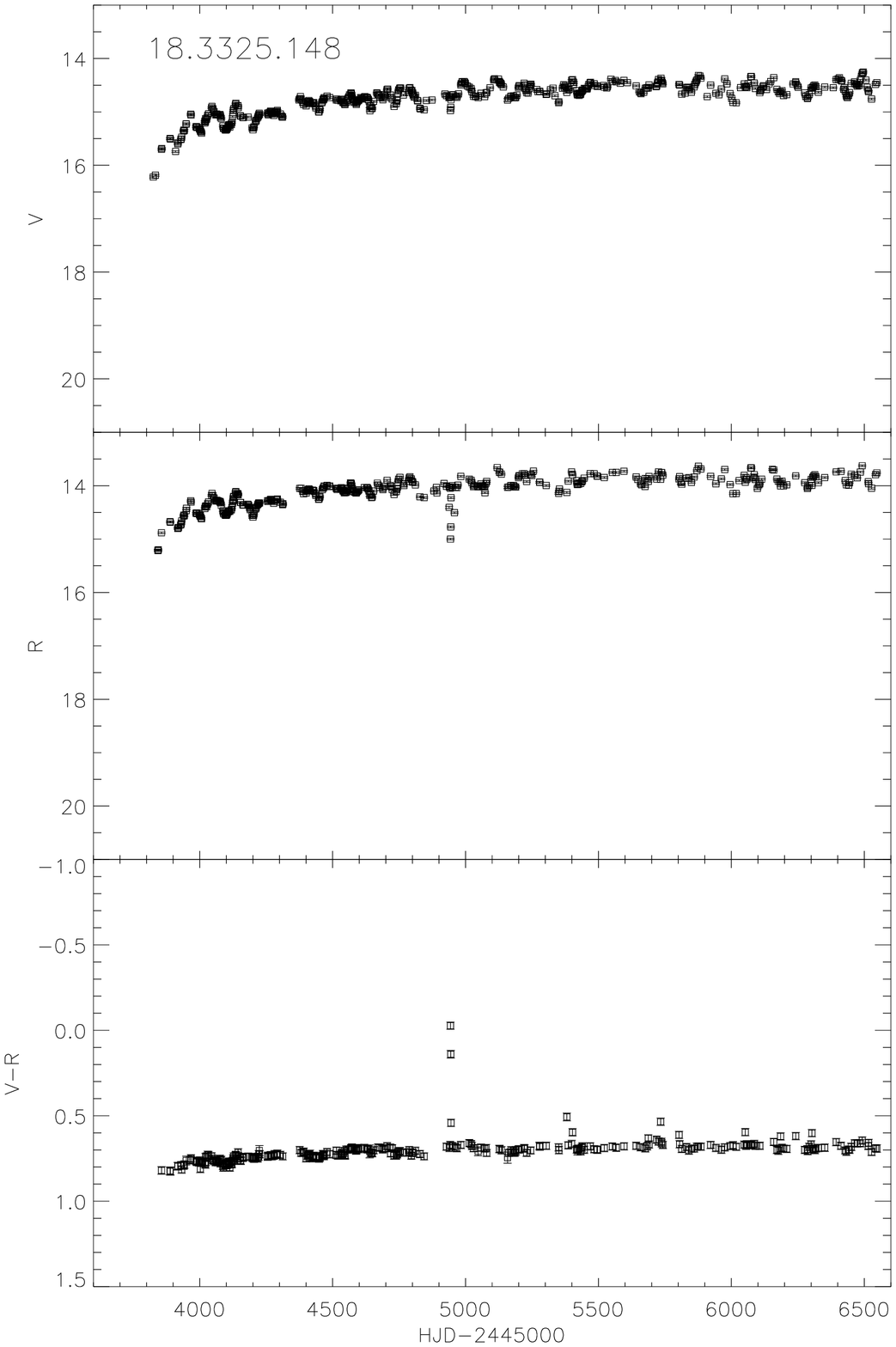}
\caption{MACHO lightcurves for confirmed RCB stars. }
%\epsscale{1.0}
\end{figure*}

\begin{figure*}
\figurenum{1d}
\epsscale{0.75}
\plottwo{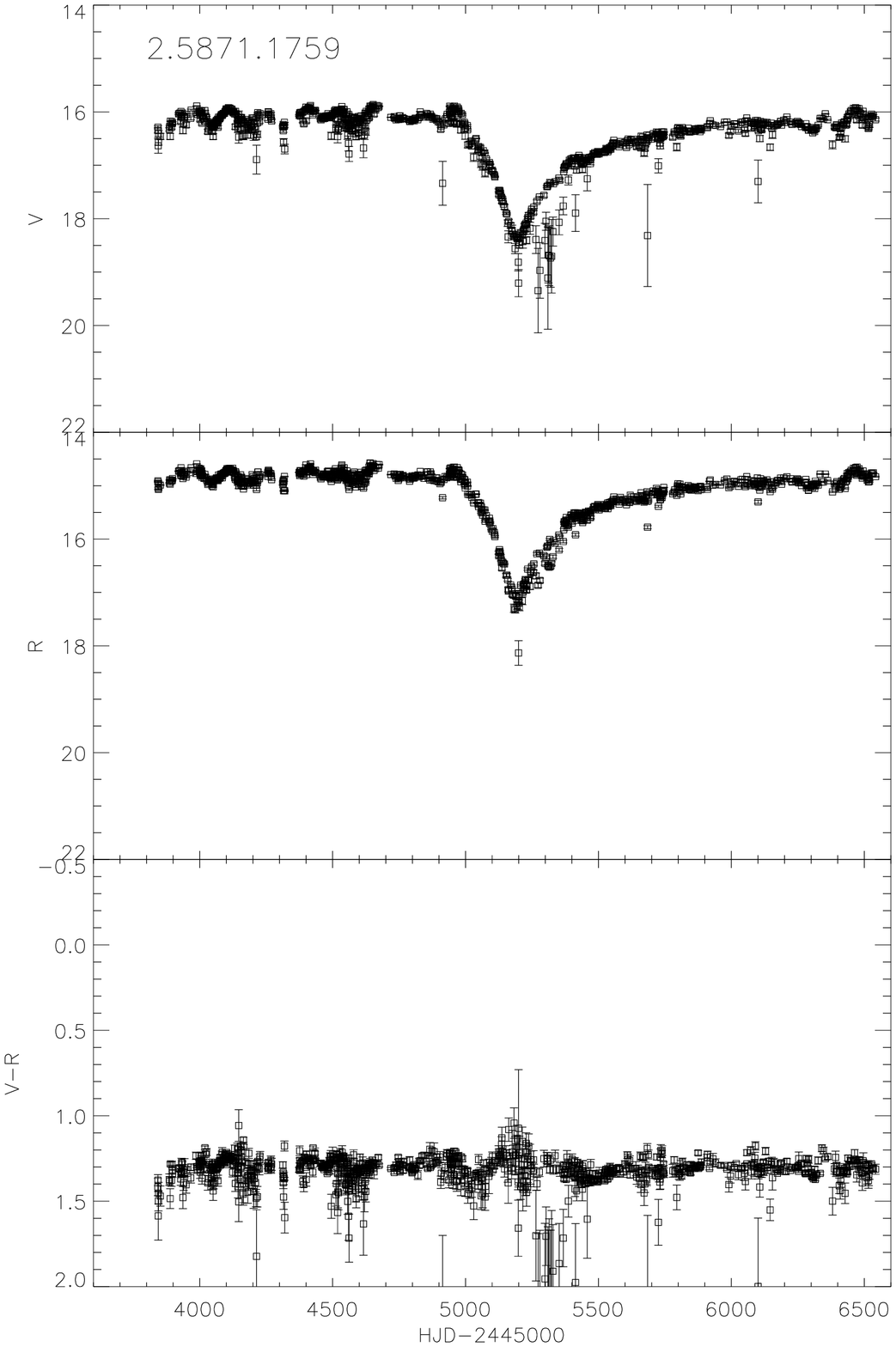}{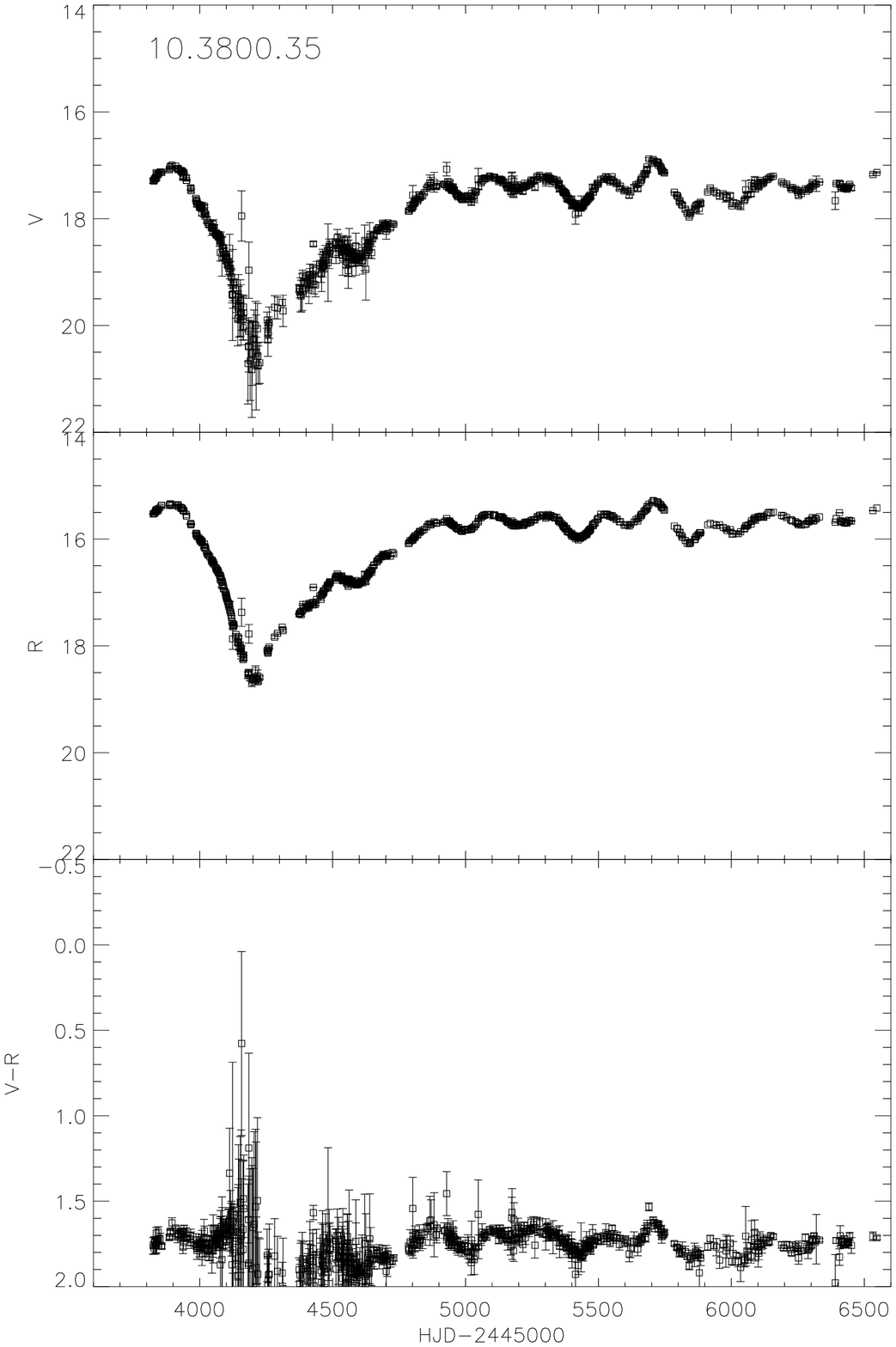}\\
\plottwo{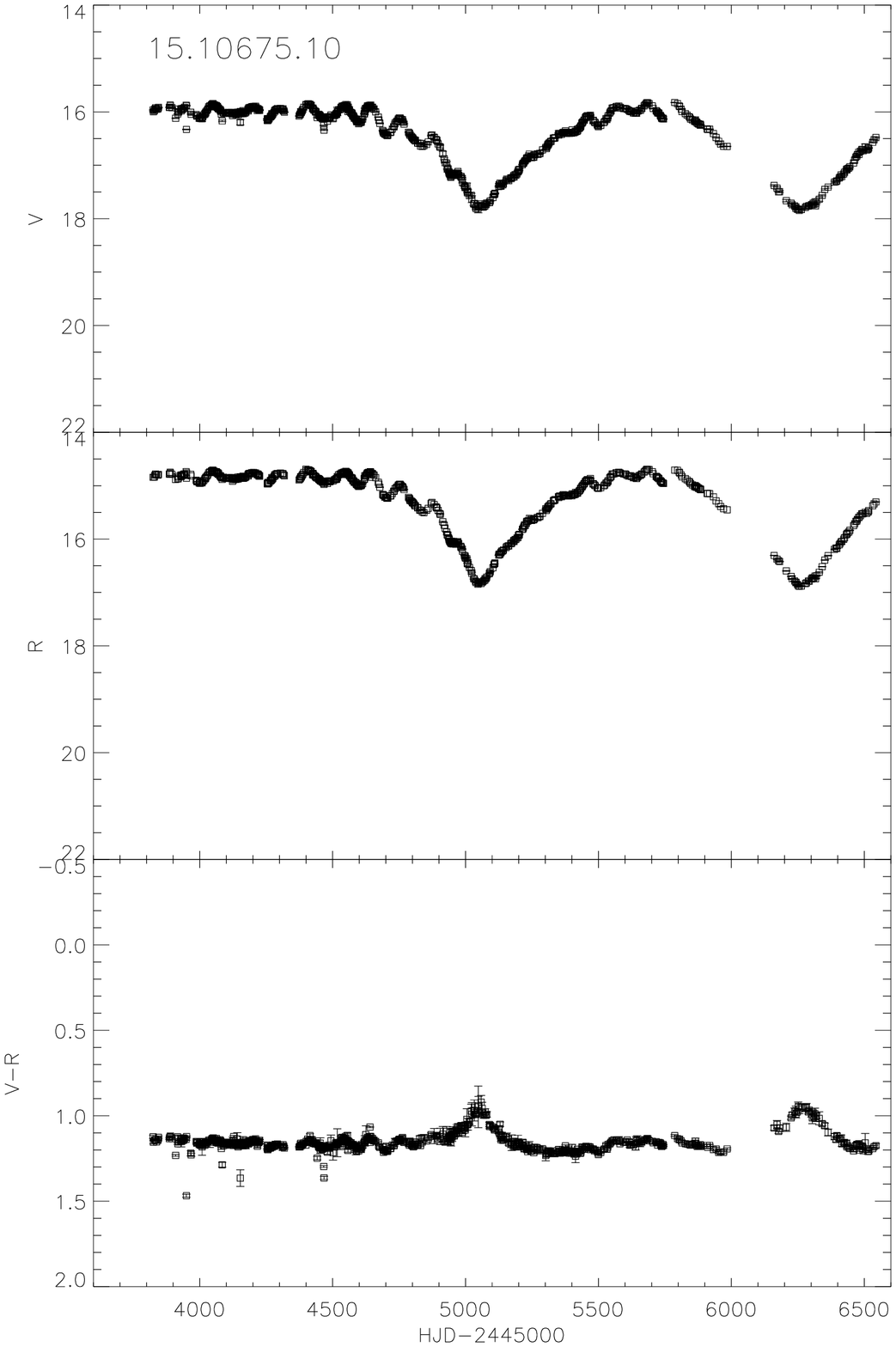}{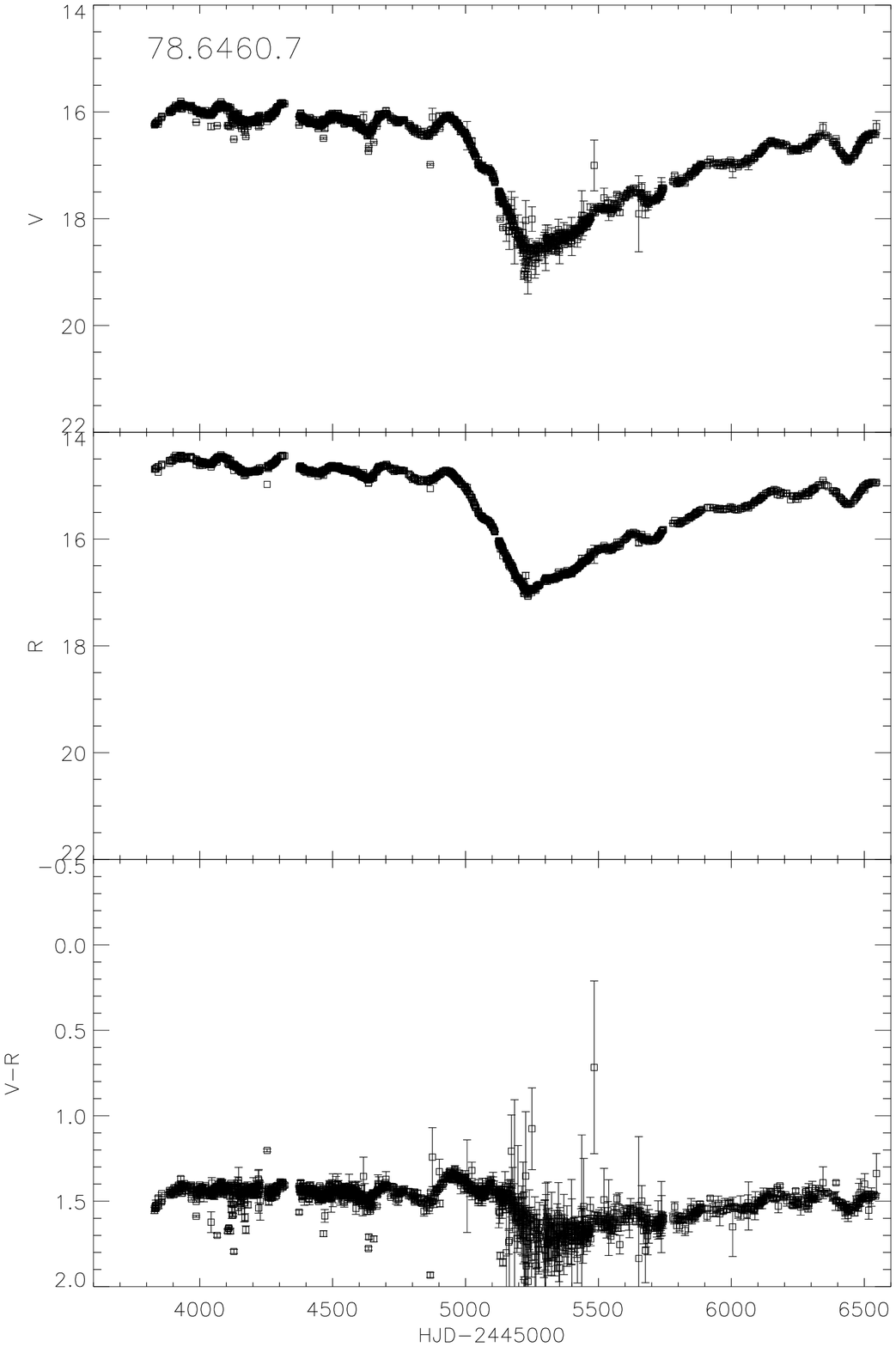}
\caption{MACHO lightcurves for DY Per type stars. }
%\epsscale{1.0}
\end{figure*}

\begin{figure*}
\figurenum{2a}
\epsscale{0.5}
\plottwo{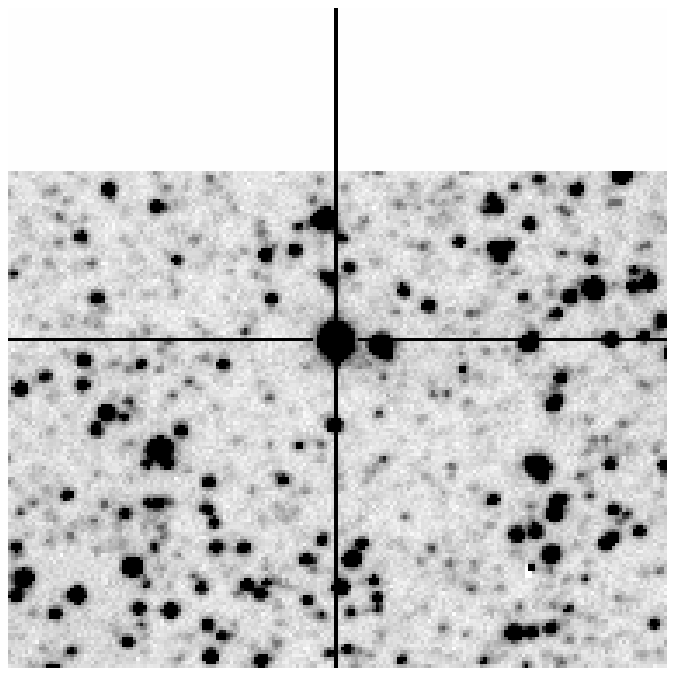}{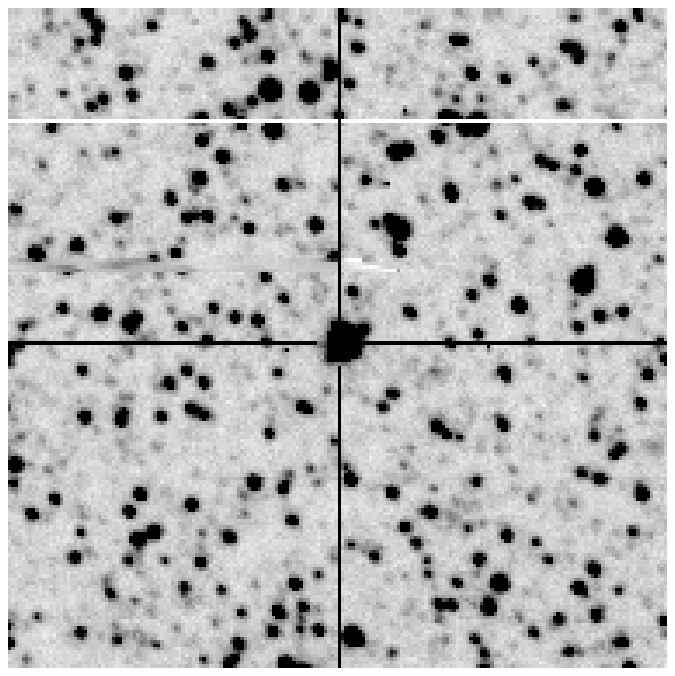}\\
\vspace{0.2cm}
\plottwo{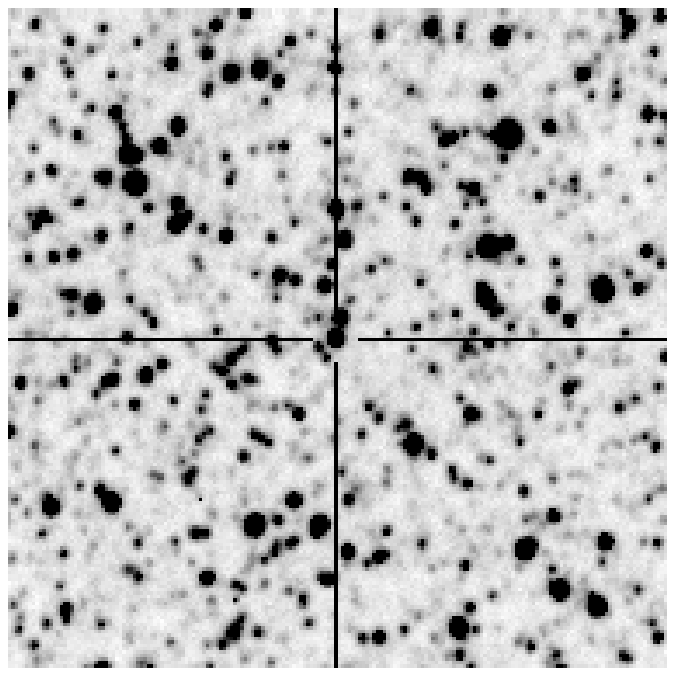}{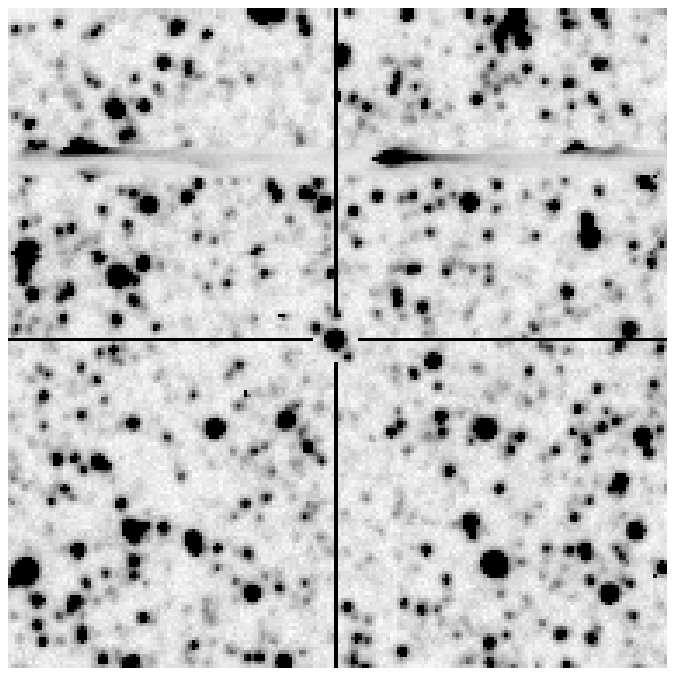}\\
\vspace{0.2cm}
\plottwo{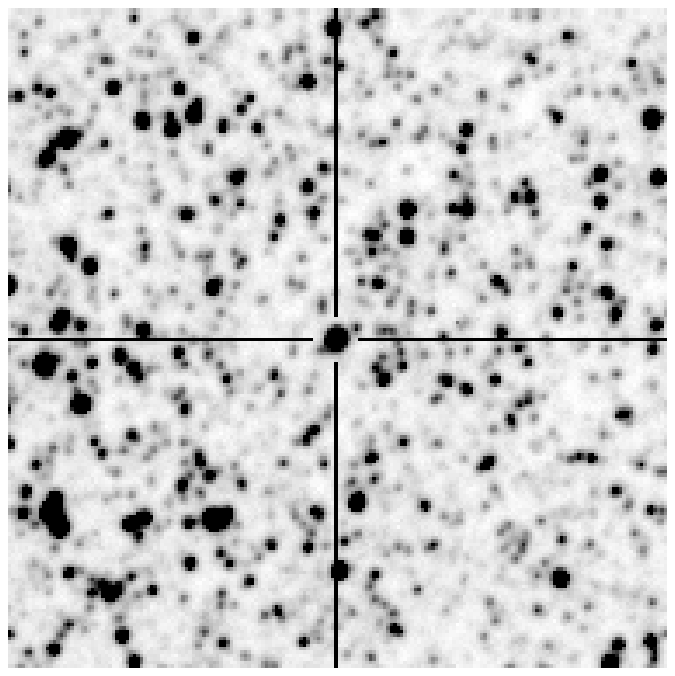}{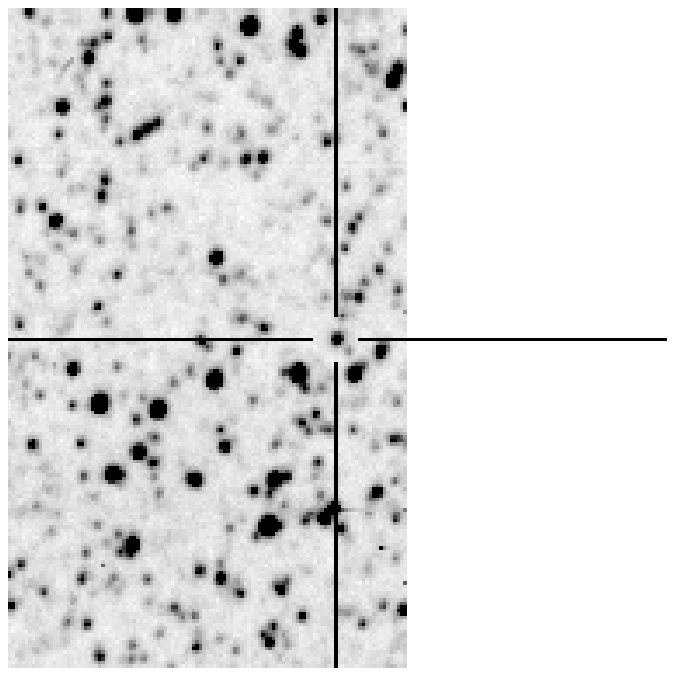}\\

\caption{Fields of confirmed RCB stars. Each field is 2' square. 
North is up and east to left.
The fields are HV 5637 (upper left), W Men (upper right), 6.6696.60 
(second row left), 12.10803.56 (second row right), 79.5743.15 (third row left),
80.6956.207 (third row right).}
%\epsscale{1.0}
\end{figure*}

\begin{figure*}
\figurenum{2b}
\epsscale{0.5}
\plottwo{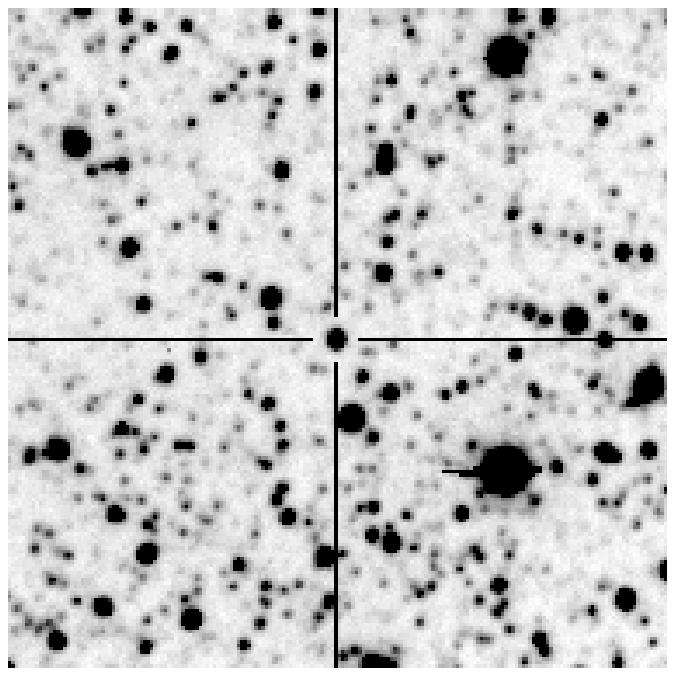}{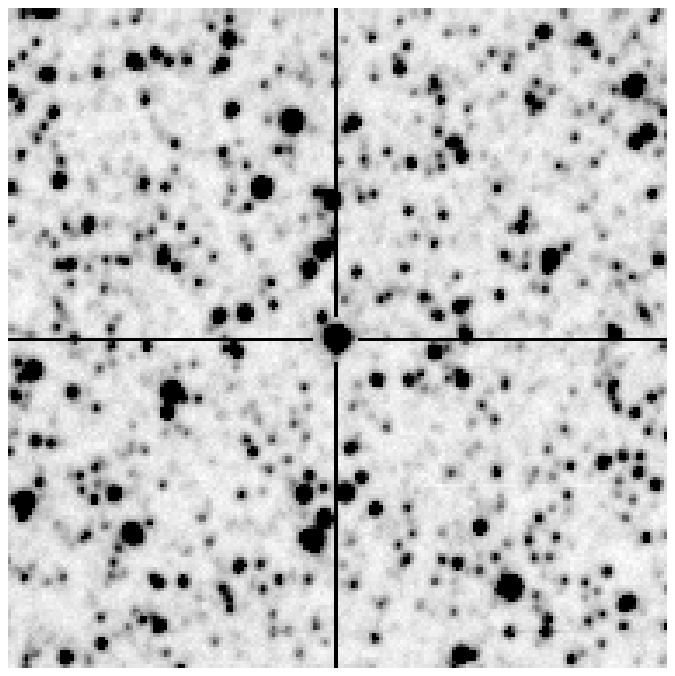}\\
\vspace{0.2cm}
\epsscale{0.5}
\plottwo{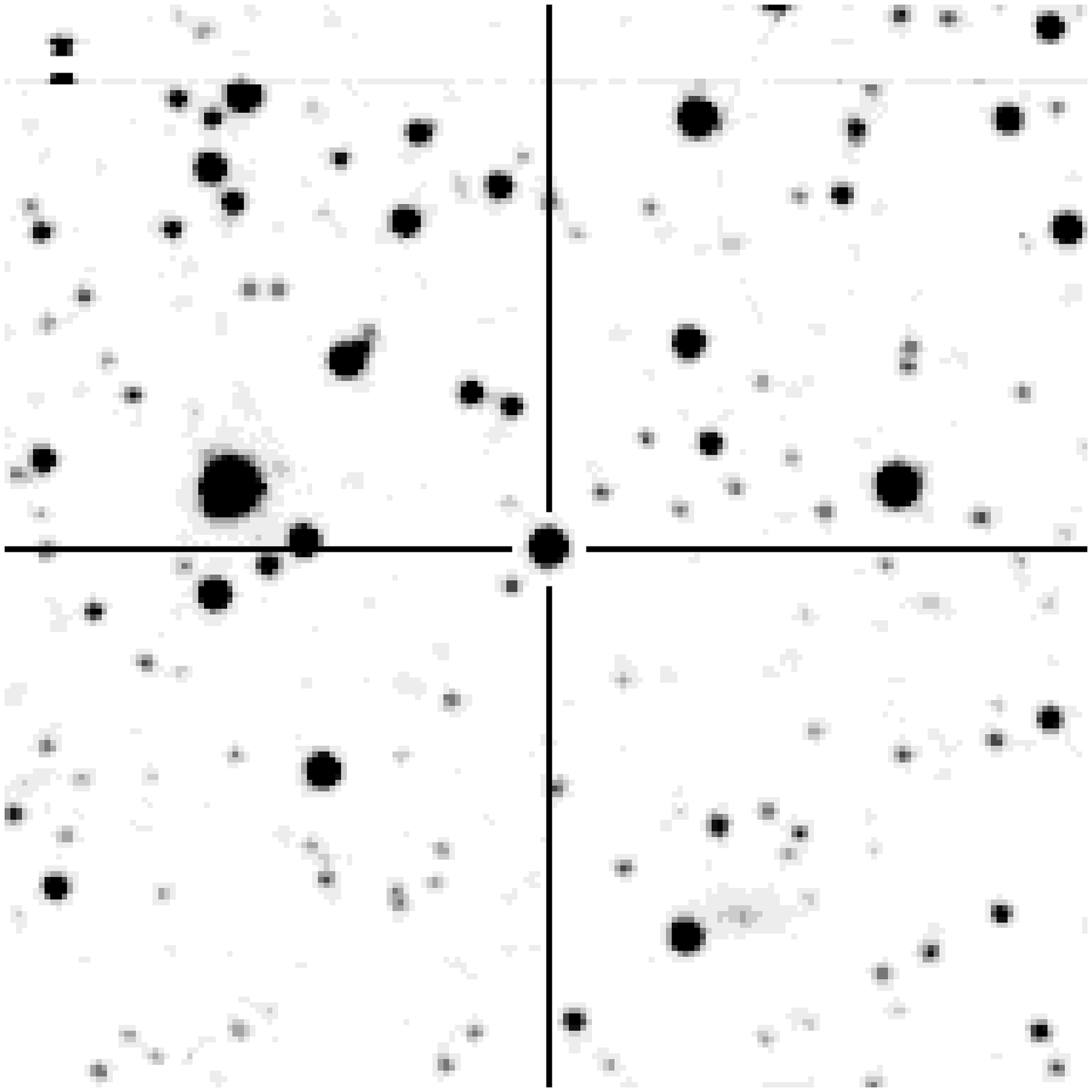}{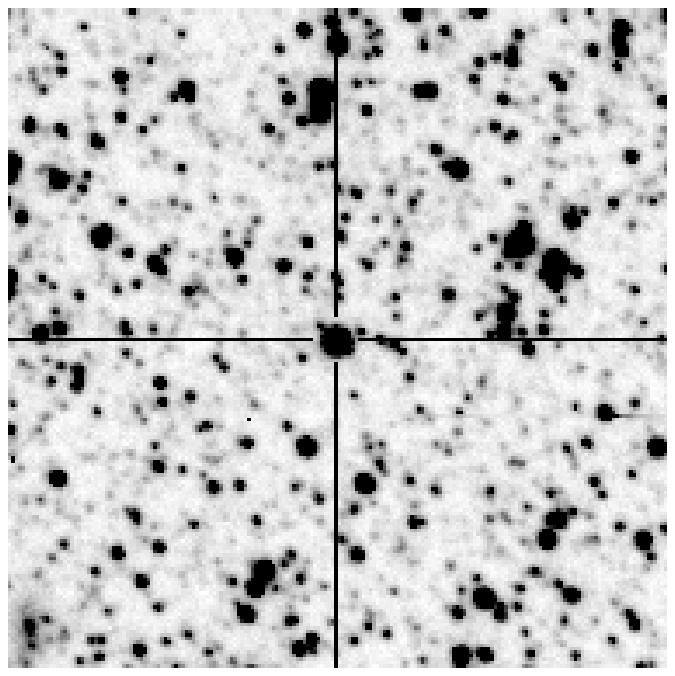}
\caption{Fields of confirmed RCB stars. Each field is 2' square. 
The fields are 80.7559.28 (upper left), 6.6575.13 (upper right), 16.5641.22 (bottom left), 18.3325.148 
(bottom right).}
%\epsscale{1.0}
\end{figure*}

\begin{figure*}
\figurenum{2c}
\hspace{1.5cm}
\epsscale{0.14}
\plotone{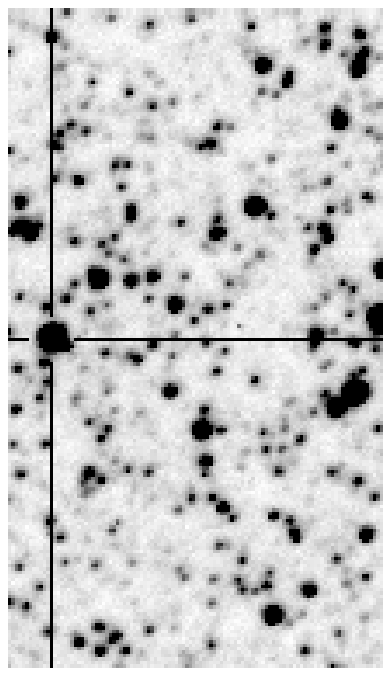}
\hspace{2.5cm}
\epsscale{0.25}
\plotone{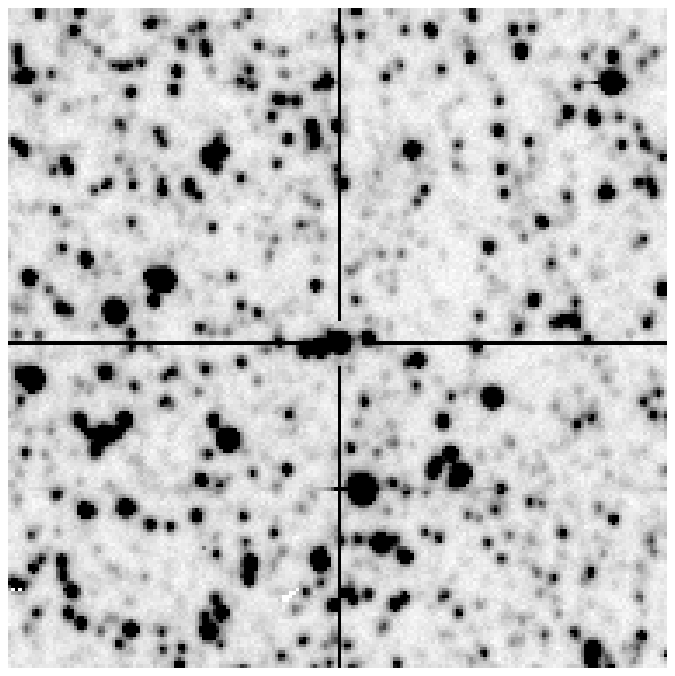}\\
\vspace{0.2cm}
\epsscale{0.5}
\plottwo{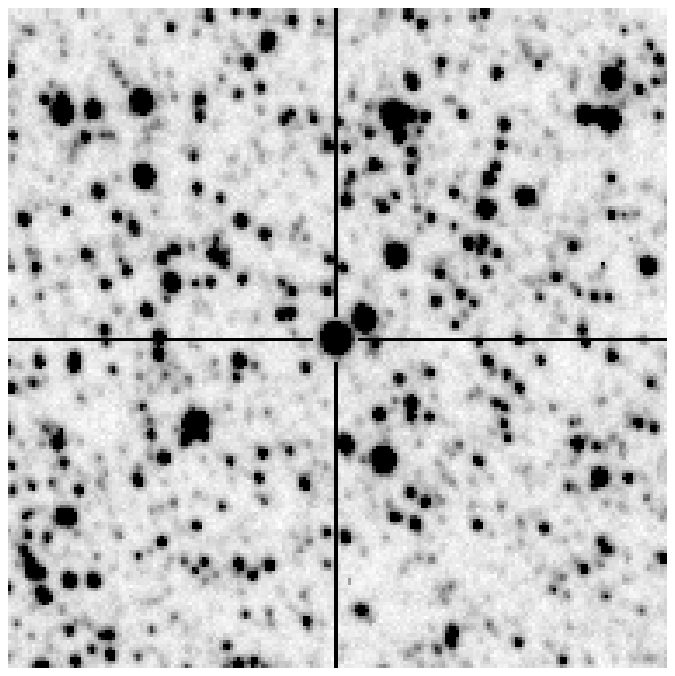}{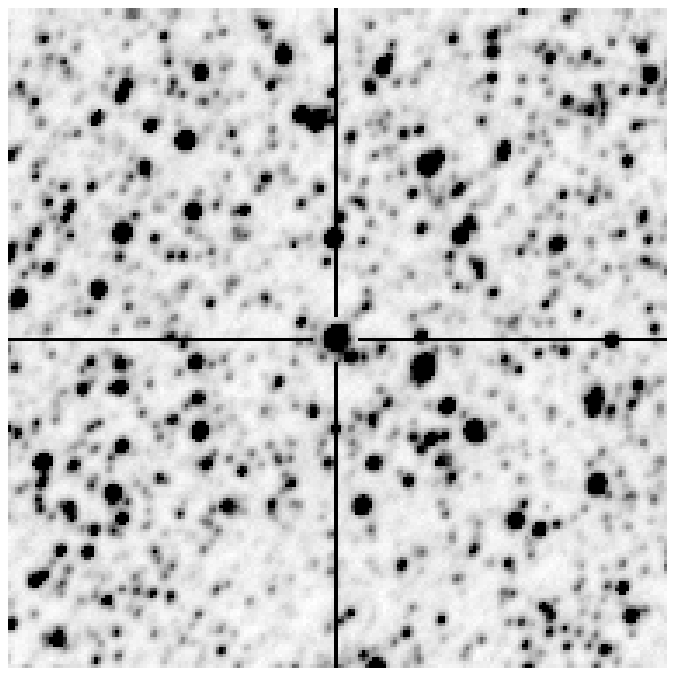}\\
\caption{Fields of DY Per type stars. Each field is 2' square. 
The fields are 2.5871.1759 (upper left), 10.3800.35 (upper right), 15.10675.10 (bottom left), and 78.6460.7  (bottom right).}
%\epsscale{1.0}
\end{figure*}

\begin{figure*}
\figurenum{3a}
\epsscale{0.5}
\plotone{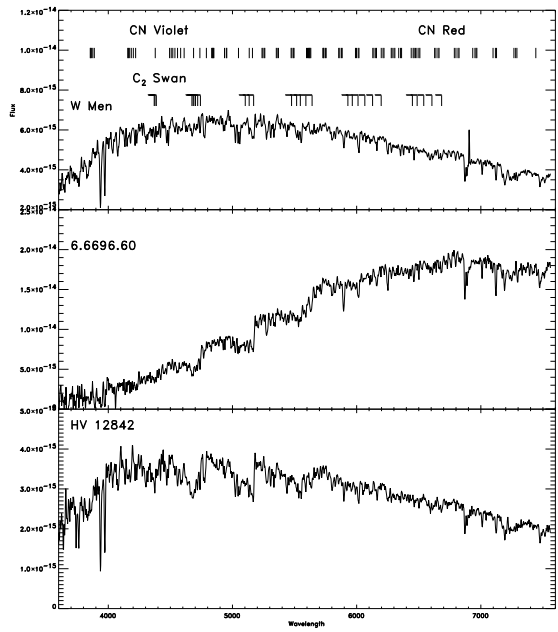}
\caption{Spectra of confirmed warm RCB stars. }
%\epsscale{1.0}
\end{figure*}

\begin{figure*}
\figurenum{3b}
\epsscale{1.0}
\plottwo{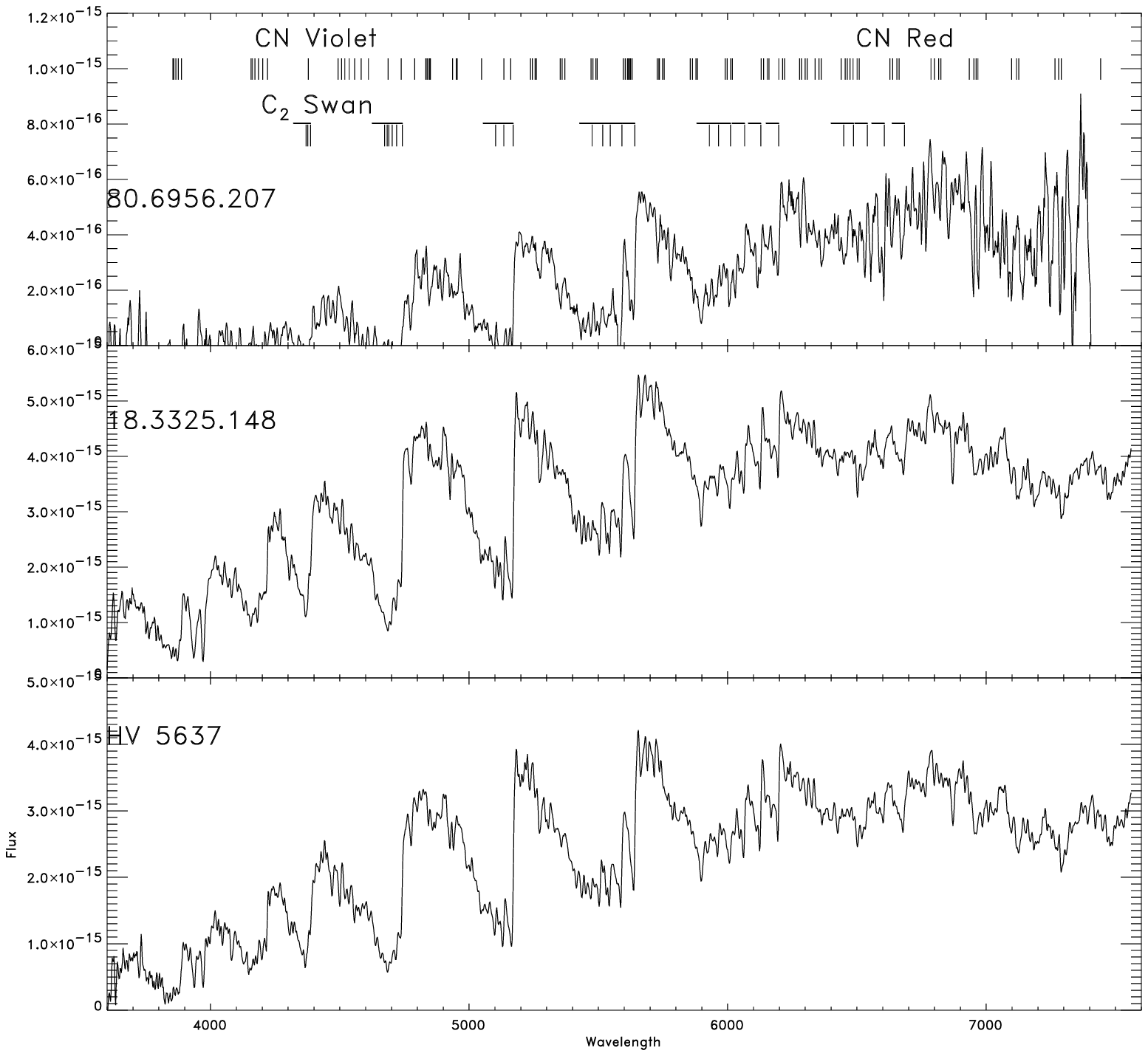}{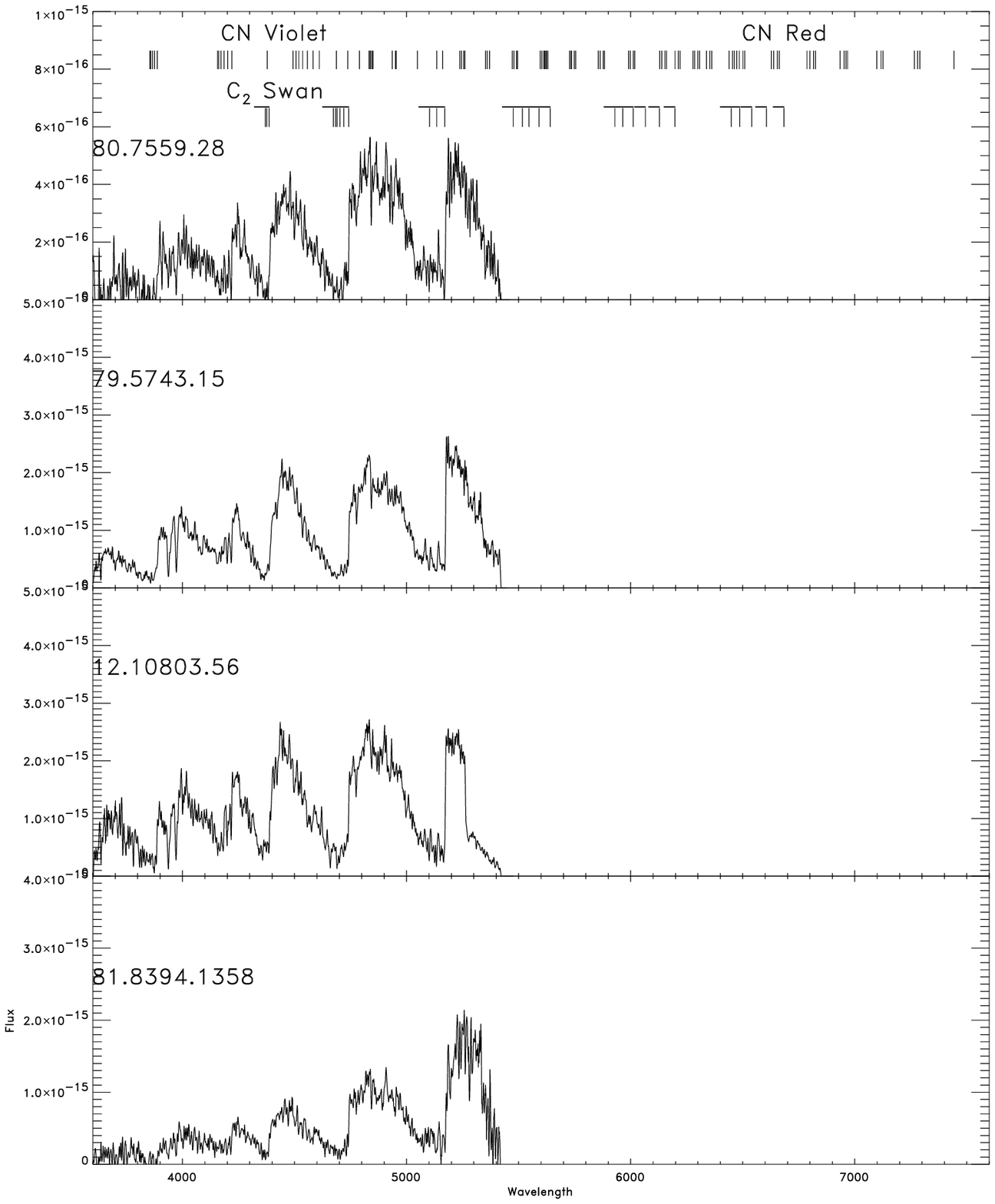}
\caption{Spectra of confirmed cool RCB stars. }
%\epsscale{1.0}
\end{figure*}

\begin{figure*}
\figurenum{3c}
\epsscale{0.5}
\plotone{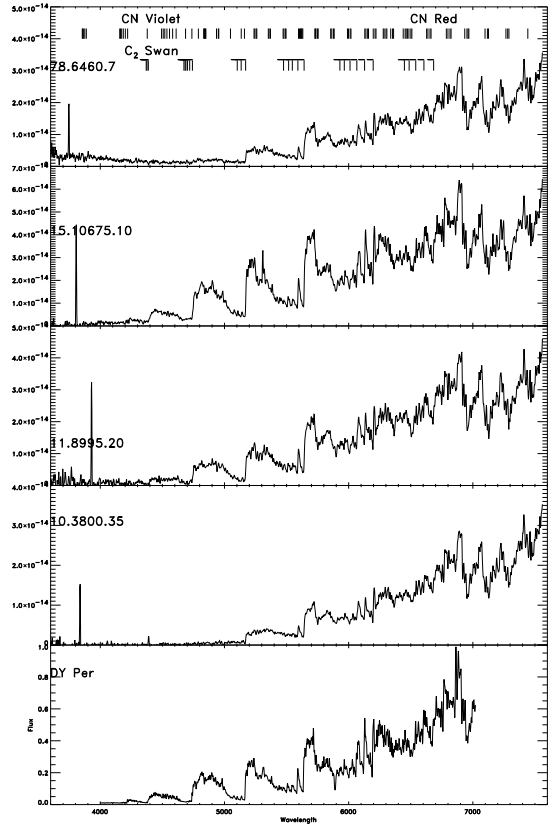}
\caption{Spectra of DY Per type stars. }
%\epsscale{1.0}
\end{figure*}

\begin{figure*}
\figurenum{4}
\epsscale{0.75}
\plotone{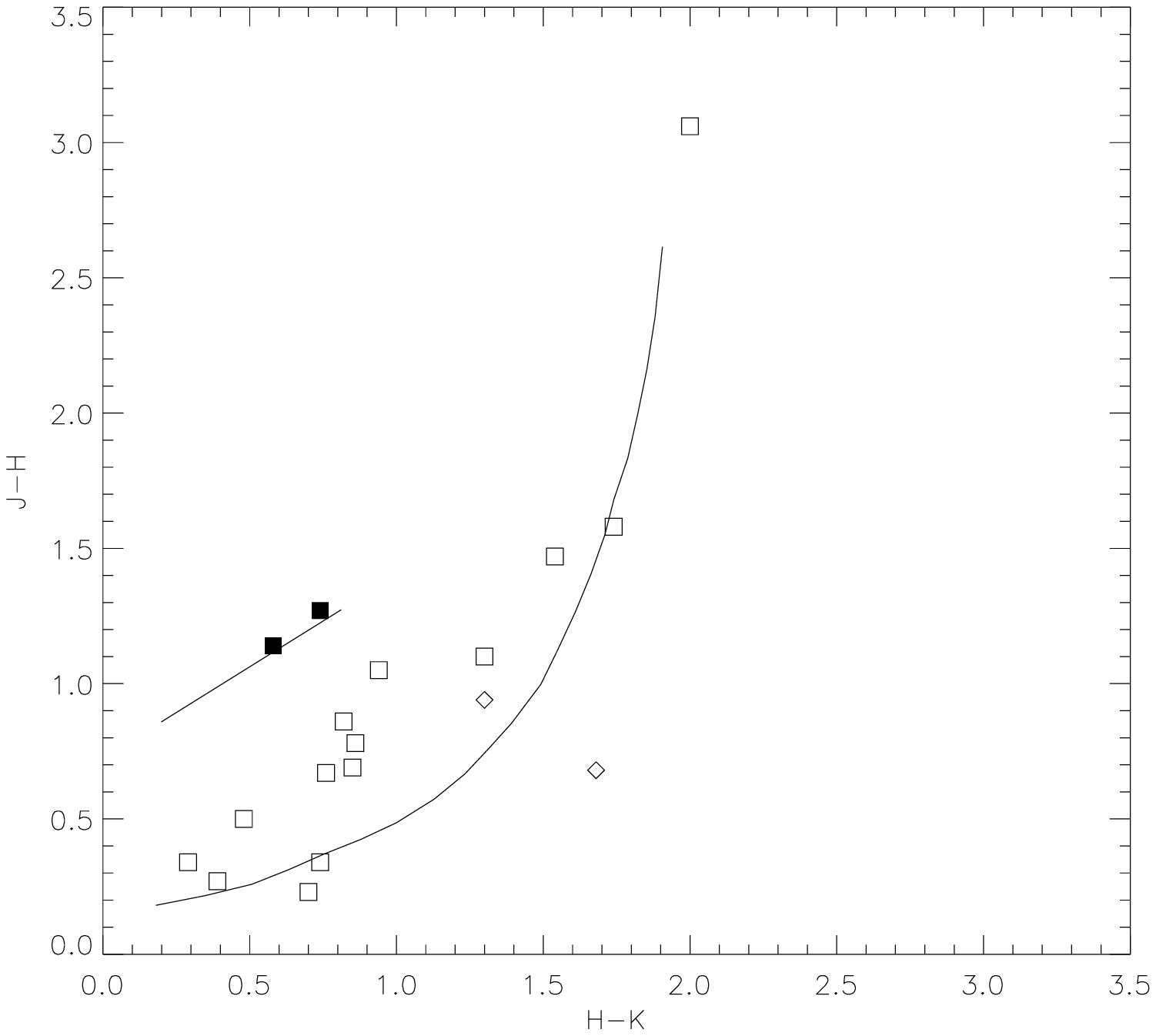}
\caption{Infrared colors for the sample are plotted for the LMC RCB (open squares) and DY Per stars 
(filled squares). 
The observations of the one hot RCB star, 11.8632.2507, are shown as open 
diamonds. Stars with multiple IR observations are plotted more than once. The upper solid line represents the colors of
LMC carbon stars (Westerlund et al. 1991). The lower line is combination of a 5500 K star and a 900 K 
dust shell blackbody ranging from all star
to all shell. The curve is taken from Feast (1997).}
%\epsscale{1.0}
\end{figure*}

\begin{figure*}
\figurenum{5}
\epsscale{0.75}
\plotone{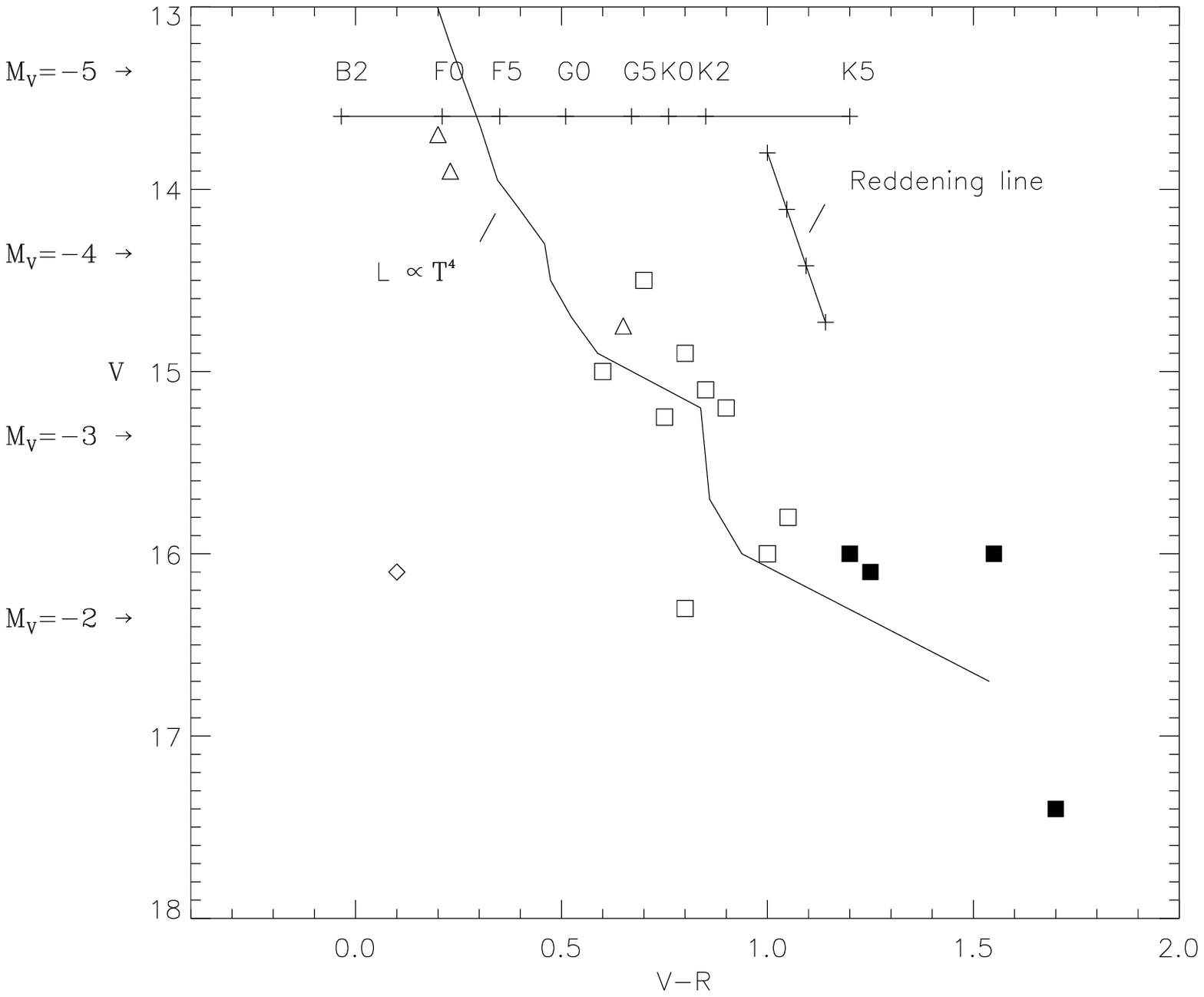}
\caption{Color-Magnitude Diagram for the new LMC RCB (open squares) and DY Per stars (filled squares). 
The three pre-MACHO RCB stars are shown as open triangles. The one hot RCB star, 11.8632.2507 is shown as a open 
diamond. The plotted reddening vector shows the effect of average Galactic reddening of E(B-V) = 0.1, 0.2, 0.3 mag.
The L$\propto~T^4$ line is explained in the text.}
%\epsscale{1.0}
\end{figure*}

\begin{figure*}
\figurenum{6}
\epsscale{1.0}
\plotone{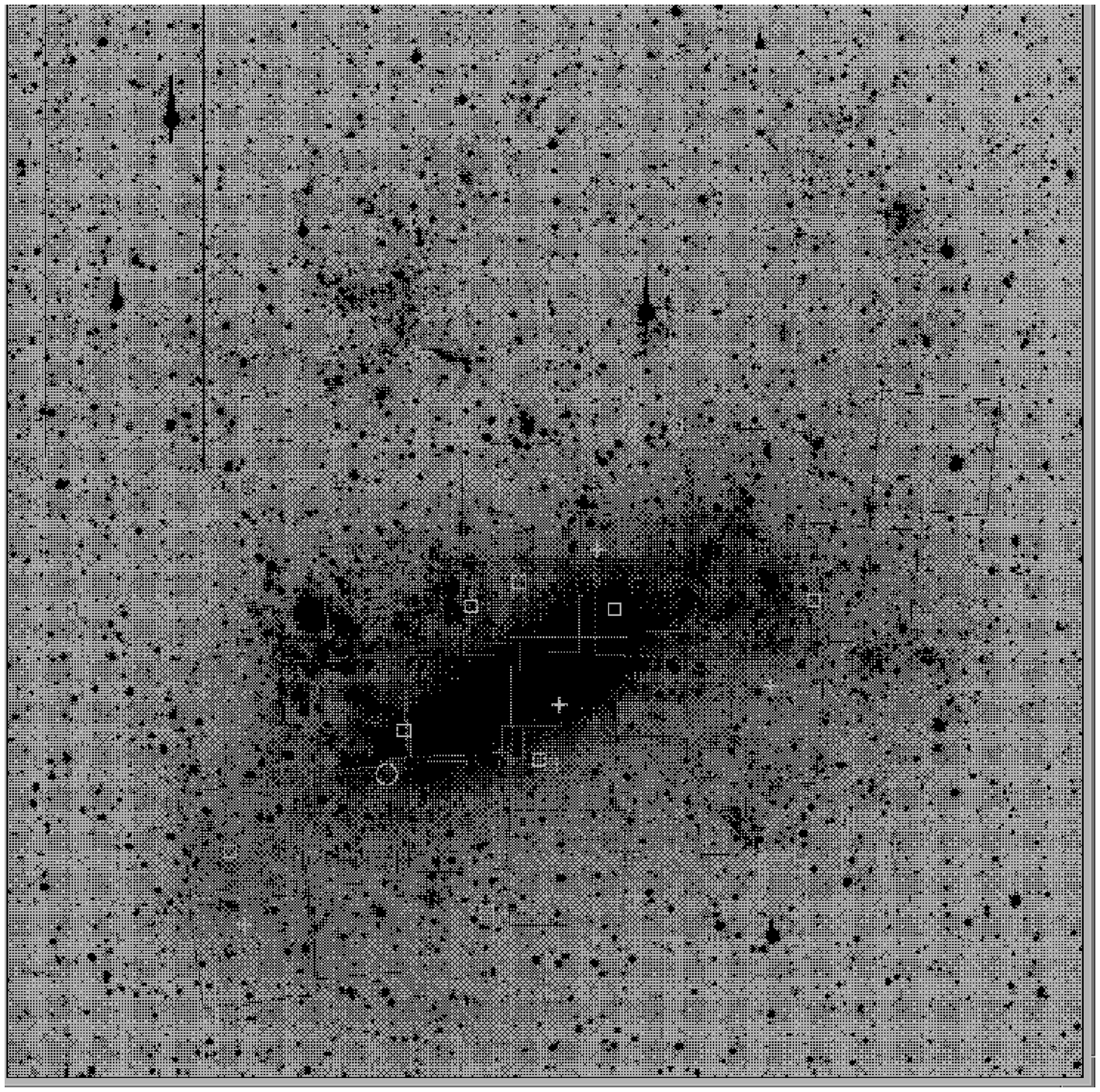}
\caption{Locations of the RCB (squares) and DY Per stars (crosses) plotted on an image of the LMC 
(Bothun \& Thompson 1988). The one hot RCB star, HV 2671 is marked with a circle. HV 12842 lies off the top of this 
image.}
\end{figure*}


\begin{references}

\reference{} Alcock, C., et al. 1992, in Robotic Telescopes in the 1990s, ASP Conf. Ser. 
No. 34, edited by A. V. Fillippenko, p. 193

\reference{} Alcock, C., et al. 1996, ApJ, 470, 583

\reference{} Alcock, C., et al. 1997a, ApJ, 486, 697

\reference{} Alcock, C., et al. 1997b, ApJ, 491, L11

\reference{} Alcock, C., et al. 1999, PASP, 111, 1539

\reference{} Alcock, C., et al. 2000, AJ, 119, 2194

\reference{} Alksnis, A. 1994, Baltic Astr., 3, 410

\reference{} Allen, D. A. 1980, Ap. Lett., 20, 131

\reference{} Asplund, M. et al. 1999, A\&A, 343, 507

\reference{} Barnbaum, C., Stone, R.P.S., \& Keenan, P.C. 1996, ApJS, 105, 419

\reference{} Bessell, M.S., \& Wood, P.R. 1983, MNRAS, 202, 31p

\reference{} Bidelman, W.P. 1948, ApJ, 107, 413

\reference{} Blanco, V.M., \& McCarthy, M.F. 1990, AJ, 100, 674

\reference{} Bothun, G.D., \& Thompson, I.B. 1988, AJ, 96, 877

\reference{} Butler, C.J. 1978, A\&AS, 32, 83

\reference{} Clayton, G. C., De Marco, O., Kilkenny, D.M., \& Pollacco, D.L. 2001, in preparation

\reference{} Clayton, G. C., \& De Marco, O. 1997, AJ, 114, 2679

\reference{} Clayton, G. C. 1996, PASP, 108, 225

\reference{} Clayton, G. C., Whitney, B. A., \& Mattei, J. 1993, PASP, 105, 832

\reference{} Cook, K. H. et al. 1995, in Astrophysical Applications of Stellar Pulsation, 
IAU Coll. No. 155, ASP Conf. Ser., 83, 221

\reference{} Cottrell, P. L., \& Lawson, W. A. 1998, PASA, 15, 179

\reference{} Cox, A.N. 2000, in Allen's Astrophysical Quantities, (Springer Verlag:New York)

\reference{} Drilling, J. S. 1986, in Hydrogen Deficient Stars and Related Objects, ed. K. 
Hunger (Dordrecht, Reidel), p. 9

\reference{} Feast, M. W. 1956, MNRAS, 116, 583

\reference{} Feast, M. W. 1972, MNRAS, 158, 11p

\reference{} Feast, M. W. 1979, Changing Trends in Variable Star Research, IAU Coll. 
No. 46, eds. F. M. Bateson, J. Smak, \& I. M. Urch (Hamilton, Univ. of Waikato), p. 
246

\reference{} Feast, M. W., Whitelock, P.A., Catchpole, R.M., \& Roberts, G. 1984, MNRAS, 211, 331

\reference{} Feast, M. W. 1997, MNRAS, 285, 339

\reference{} Glass, I. S. 1988, IAU Circ. No. 4572

\reference{} Glass, I. S., Lawson, W. A., \& Laney, C. D. 1994, MNRAS, 270, 347

\reference{} Goldsmith, M.J., Evans, A., Albinson, J.S., \& Bode, M.F. 1990, MNRAS, 245, 119

\reference{} Gonzalez, G. et al. 1998, ApJS, 114, 133

\reference{} Hart, J. et al. 1996, PASP, 108, 220

\reference{} Hodge, P.W., \& Wright, F.W. 1967, The Large Magellanic Cloud,
Smithsonian Publication 4699 (Washington; Smithsonian Press)

\reference{} Hodge, P.W., \& Wright, F.W. 1969, ApJS, 17, 467

\reference{} Hughes, S. M.G. 1989, AJ, 97, 1634

\reference{} Iben, I. Jr., \& Tutukov, A.V. 1985, \apjs, 58, 661

\reference{} Iben, I., Tutukov, A. V., \& Yungelson, L. R. 1996a, ApJ, 456, 750

\reference{} Iben, I., Tutukov, A. V., \& Yungelson, L. R. 1996b, in Hydrogen Deficient Stars, ASP Conf. Ser., 96, 409

\reference{} Keenan, P.C., \& Barnbaum, C.B. 1997, PASP, 109, 969

\reference{} Kerber, F. et al. 1999, A\&A, 344, L79

\reference{} Kilkenny, D., and Marang, F. 1989, MNRAS, 238, 1p

\reference{} Kim, S., Staveley-Smith, L., Dopita, M.A., Freeman, K.C., Sault, R.J., Kesteven, M.J., \& McConnell, D. 1998,
ApJ, 503, 674

\reference{} Kurochkin, N.E. 1992, Sov. Astron. Lett. 18, 410

\reference{} Lawson, W. A., Cottrell, P. L., Kilmartin, P. M., \& Gilmore, A. C. 1990, 
MNRAS, 247, 91

\reference{} Lawson, W. A., and Cottrell, P. L. 1989, MNRAS, 240, 689

\reference{} Lawson, W. A., \& Cottrell, P. L. 1990a, Observatory, 110, 132

\reference{} Lawson, W. A., \& Cottrell, P. L. 1990b, Confrontation Between Stellar 
Pulsation and Evolution, ASP Conf. Ser., 11, 566

\reference{} Lawson, W. A., Cottrell, P. L., \& Pollard, K.R. 1991, in The Magellanic Clouds, IAU Symp. No. 
148, eds., R. Haynes and D. Milne, (Kluwer: Dordrecht), p. 351

\reference{} Lloyd Evans, T. 1997, MNRAS, 286, 839

\reference{} Luyten, W.J. 1927, Harvard Bull., 846, 33

\reference{} Mattei, J. A., Waagen, E. O., \& Foster, G. 1991, AAVSO Monograph No. 4

\reference{} Milone, L. A. 1975, IBVS No. 989

\reference{} Morgan, D.H., Nandy, K., \& Rao, N.K. 1986, in Hydrogen Deficient Stars and Related Objects, 
ed. K. Hunger (Dordrecht, Reidel), p. 225

\reference{} Morgan, D.H. 1994, A\&AS, 103, 235

\reference{} Nelson, C.A., Cook, K.H., Popowski, P., \& Alves, D.R. 2000, AJ, 119, 1205

\reference{} Oestreicher, M. O., Gochermann, J. \& Schmidt--Kaler, T. 1995, A\&AS, 112, 495

\reference{} Oestreicher, M. O., \& Schmidt--Kaler, T. 1996, A\&AS, 117, 303

\reference{} Payne-Gaposchkin, C.E. \& Gaposchkin, S. 1938, Variable Stars, Harvard University Monograph No. 5
(Cabridge: Harvard College Observatory)

\reference{} Payne-Gaposchkin, C. E. 1971, Smithsonian Contr., No. 13, 1

%\reference{} Pollard, K. R., Cottrell, P. L., \& Lawson, W. A. 1994, MNRAS, 268, 544

\reference{}  Richer, H.B. 1981, ApJ, 243, 744

\reference{} Roberts, D.H., Lehar, J. and Dreher, J.W. 1987, AJ, 93, 968

\reference{} Rodgers, A. W. 1970, Observatory, 90, 197

\reference{} Sanduleak, N. \& Philip, A.G.D.  1977, Warner \& Swasey Obs., Pub. No. 2, 105

\reference{} Schecter, P.L. , Mateo, M., \& Saha, A. 1993, PASP, 105, 1342

\reference{} Sch\"{o}nberner, D. 1986, in Hydrogen Deficient Stars and Related Objects, 
ed. K. Hunger (Dordrecht, Reidel), p. 221

\reference{} Schwering, P. B. W. 1989, A\&AS, 79, 105

\reference{} Schwering, P. B. W., \& Israel, F. P. 1991, A\&A, 246, 231

\reference{} Stubbs, C. W., et al. 1993 in Charged-Coupled Devices and Solid State 
Optical Sensors III, edited by M. Blouke, Proc of the SPIE, 1900, 192

\reference{} Trams, N.R. et al. 1999, A\&A, 346, 843

\reference{} Trimble, V, \& Kundu, A. 1997, PASP, 109, 1089

\reference{} Van Loon, J.T., Groenewegen, M.A.T., De Koter, A., Trams, N.R., Waters, L.B.F.M., Zijlstra, A.A., 
Whitelock, P.A., Loup, C. 1999, A\&A, 351, 559

\reference{} Westerlund, B.E., Olander, N., Richer, H.B., \& Crabtree, D.R. 1978, A\&AS, 31, 61

\reference{} Westerlund, B.E., Azzopardi, M., Breysacher, J., \& Rebeirot, E. 1991, A\&AS, 91, 425

\reference{} Webbink, R. F. 1984, ApJ, 277, 355

\reference{} Weiss, A. 1987, A\&A, 185, 178

\reference{} Wood, P.R., \& Cohen, M. 2001, in Post-AGB objects as a phase of stellar evolution,
eds. R. Szczerba \& S.K. Gorny, in press

\reference{} Wright, F.W., \& Hodge, P.W. 1971, AJ, 76, 1003

\reference{} Yuin, C. 1948, ApJ, 107, 413

\end{references}
\end{document}